\documentclass[aps,twocolumn,amsmath,amssymb,superscriptaddress,showpacs]{revtex4-2}%
\usepackage{graphicx}
\usepackage{bm}
\usepackage{float}
\usepackage{subfig}

\usepackage{xcolor}
\usepackage{braket}
\usepackage{tabularx}
\usepackage{comment}
\usepackage{afterpage}
\usepackage{placeins}
\usepackage{booktabs}
\usepackage{multirow}
\usepackage{array}
\usepackage{setspace}
\graphicspath{{Figs/}}
\usepackage{hhline}
\usepackage{xfrac}
\usepackage{mathtools}
\usepackage{listings}
\usepackage[normalem]{ulem}
\usepackage{titlesec}
\usepackage{amsfonts}
\usepackage[version=4]{mhchem}

\usepackage{epstopdf}
\catcode`@11
\def\seceqaa{\@addtoreset{equation}{section}
principles\def\theequation{A\arabic{equation}}}
\def\seceqbb{\@addtoreset{equation}{section}
\def\theequation{B\arabic{equation}}}
\def\seceqcc{\@addtoreset{equation}{section}
\def\theequation{C\arabic{equation}}}
\def\seceqdd{\@addtoreset{equation}{section}
\def\theequation{D\arabic{equation}}}
\def\seceqee{\@addtoreset{equation}{section}
\def\theequation{E\arabic{equation}}}
\def\seceqff{\@addtoreset{equation}{section}
\def\theequation{F\arabic{equation}}}
\def\seceqgg{\@addtoreset{equation}{section}
\def\theequation{G\arabic{equation}}}
\def\seceqhh{\@addtoreset{equation}{section}
\def\theequation{H\arabic{equation}}}
\catcode`@11

\newcommand{\mz}[1]{{#1}}

\begin{document}
\title{Relation between chiral anomaly and electric transport in $1D$ Dirac semimetal}
\author{Mustafa Bohra}
\address{ \scriptsize Department of Physics, Ariel University, Ariel-40700, Israel}

\author{ Mikhail Zubkov}
\address{ \scriptsize Department of Physics, Ariel University, Ariel-40700, Israel}

\date{\today}

\begin{abstract}

\noindent We investigate the interplay of chiral anomaly and dissipation in one - dimensional Dirac semimetal. For definiteness we consider the Su-Schrieffer–Heeger (SSH) model, which on the language of lattice field theory represents $1$ D Wilson fermions. We employ the non-equilibrium Keldysh Green function formalism, and  calculate the chiral imbalance and electric conductivity in the presence of energy dissipation, revealing how these observables are connected to the chiral anomaly. By systematically incorporating dissipation effects into the Keldysh framework, we demonstrate how the anomaly-induced contributions manifest in both axial charge density and electric current. 
\end{abstract}

\maketitle

\section{Introduction}
The study of non-equilibrium quantum systems has emerged as a cornerstone of modern condensed matter physics \cite{Onoda2006}, particularly for understanding transport phenomena, topological materials, and quantum anomalies. Among these, the chiral anomaly, which is a quantum mechanical phenomenon in which the conservation laws of chiral charge break down under external electromagnetic fields, has attracted significant attention because of its profound implications in Dirac and Weyl semimetals, as well as high-energy physics.

The Su–Schrieffer–Heeger (SSH) model is  known for its role in understanding electronic properties of materials like polyacetylene \cite{PhysRevLett.42.1698, PhysRevB.108.195103,PhysRevA.100.012112, PhysRevB.96.125418}. This model is part of a class of models known as tight-binding models, which focus on how electrons move between fixed points (or lattice sites) in a material.

The Schwinger - Keldysh technique, originally formulated by Leonid V. Keldysh \cite{Keldysh64,keldysh2024diagram} in 1964 (and independently by J.Schwinger \cite{Schwinger61}), is a non-equilibrium Green’s function  approach that extends quantum many-body theory to systems driven out of equilibrium. By introducing a contour in complex plane of time, the method provides a unified way to compute time-ordered, anti-time-ordered, correlation functions, in the presence of interactions, and external fields. \mz{We use the Keldysh formalism in the ''upper-triangle'' representation \cite{Banerjee2020obs,Banerjee2022snd}   (see also the similar construction in \cite{Arseev2015} ). More details on this technique may be found in 
  \cite{Kamenev2005course,KamenevBook}.
Such definition allows us to construct the path-integral representation for the Keldysh formalism. }

\mz{Keldysh formalism was used both in condensed matter physics and in high energy physics \cite{KB62,Baym62,Schwinger61}. It is reduced to the  Matsubara technique in case of thermal equilibrium  \cite{Matsubara55,BdD59,Gaudin60,AGD63}.As in conventional quantum field theory there are two formalisms in Keldysh technique: path integral formalism   \cite{KamenevBook} and the operator formalism \cite{Langreth76,Danielewicz84,CSHY85,RS86,Berges04,HJ98,Rammer07}. Although both versions are similar to  the  Matsubara technique, the specific difficulty is present in Keldysh path integral technique related to the turning points of the Keldysh contour in complex plane of time. Naive gluing of field variables in continuum theory at the turning points results in wrong expressions for the  Green functions even for the non - interacting systems. However, the lattice regularization resolves this problem restoring the correct expressions  \cite{Kamenev2}. It is worth mentioning that the operator formalism \cite{LP,Mahan} gives the correct answers directly. }

\mz{Perturbation theory  \cite{Bonitz00} including  Schwinger-Dyson equations \cite{Landau56,LW60,Luttinger60} were used widely in quantum kinetic theory  \cite{Baym62}. In particular, this allows to derive the  Bogoliubov-Born-Green-Kirkwood-Yvons (BBGKY) sequence of equations \cite{Cercignani88} directly from the miscroscopic dynamics. Conventional  kinetic equations appear as a result of truncation from the BBGKY hierarchy \cite{CCNS71}. This has been applied, in particular, to the description of superconductivity \cite{AG75,LO75,Bonitz00,Langreth76,Danielewicz84,CSHY85,RS86,Berges04,HJ98,Rammer07}.
Among applications to the high energy physics we would like to notice  scattering in quark matter \cite{CGC}, hydrodynamics \cite{hydr}, and  cosmology  \cite{Akh}.}

\mz{Wigner distribution has been used in the framework of Keldysh technique widely  \cite{Langreth76,Danielewicz84,CSHY85,RS86,Berges04,HJ98,Rammer07,Polkovnikov:2009ys}. More involved aspects of Wigner - Weyl calculus were developed within the equilibrium quantum field theory \cite{ZW2019,ZZ2022,Khaidukov2017,Zubkov2017,SuleymanovZubkov2020,ZA2023,Z2024,XZ2024}. Even earlier  Wigner-Weyl calculus was adopted to Keldysh technique in  \cite{Shitade} (see also references therein). Later these constructions were repeated in different context in \cite{Banerjee2020obs,Banerjee2022snd}   and independently  in \cite{Mokrousov}. }

The chiral anomaly constitutes a fundamental aspect of quantum field theory, characterized by the non-conservation of the chiral current in the quantum regime \cite{ZUMINO1984477}, despite its exact conservation within the corresponding classical framework. It arises when a chiral symmetry of the classical Lagrangian is broken by quantum mechanical effects. Moreover, the chiral anomaly is intrinsically linked to topological properties of the underlying gauge field configurations, with the net imbalance between left- and right-handed fermionic states being governed by topological invariants. Chiral imbalance—an unequal population of left- and right-handed fermions—arises in systems where the chiral anomaly or other symmetry-breaking mechanisms are present, notably in high-energy environments, as well as in condensed matter realizations like Weyl and Dirac semimetals subjected to parallel electric and magnetic fields \cite{PhysRevB.108.195103, PhysRevB.27.6083, PhysRevLett.64.1812, Lohse2016, Nakajima2016}. The first observation of chiral anomaly in condensed matter systems has been reported in \cite{bevan1997momentum} about 30 years ago.

Within the classical Drude framework \cite{Drude1900}, electrical conductivity arises from the motion of charge carriers---typically electrons---subject to an external electric field and undergoing momentum-relaxing scattering processes. In this picture, the conductivity is expressed as 
\[
\sigma = \frac{ne^{2}\tau}{m},
\]
where \(n\) is the carrier density, \(e\) the electronic charge, \(m\) the effective mass, and \(\tau\) the relaxation time determined by scattering events. The Drude model captures the essential low-frequency, or ``Drude'' mode, response of metals and semimetals, providing a baseline description against which deviations due to quantum, many-body, or topological effects can be identified.

In \(1+1\)-dimensional systems hosting chiral fermions, a finite chiral imbalance can strongly influence electrical conductivity through anomaly-induced transport processes. When parallel electric and magnetic fields---or their lower-dimensional analogs---are applied, the chiral anomaly drives a net transfer of charge between left- and right-moving fermionic modes, leading to an excess population of one chirality over the other. This imbalance modifies the current--field relationship by introducing an additional contribution to the conductivity, which depends on the rate of chiral charge relaxation in the system. As a result, the electrical response is governed not only by conventional scattering mechanisms but also by the nonequilibrium dynamics of chiral charge \cite{PhysRevB.27.6083, PhysRevLett.64.1812, Lohse2016, Nakajima2016}.

The interplay between chiral imbalance, electrical transport, and the chiral anomaly forms a key mechanism governing nontrivial current responses in systems with relativistic-like fermionic excitations. In the presence of an external electric field, the anomaly induces a spectral flow between left- and right-moving modes, breaking the separate conservation of their respective currents. This continuous redistribution of carriers generates a nonequilibrium chiral imbalance, which in turn alters the longitudinal electrical conductivity by introducing an anomaly-driven contribution in addition to the conventional Drude response. The magnitude of this contribution is controlled by both the anomaly coefficient and the relaxation dynamics of chiral charge, thereby linking quantum field–theoretic symmetry breaking to measurable transport properties.

The methodology adopted in the present research represents an application of Keldysh field theory to the calculation of axial charge density and electric conductivity in the presence of dissipation modeled by the finite dissipation rate. 

The purpose of the present work is to compare response of the chiral density to electric field with electric conductivity, for the typical 1D Dirac semimetal. For this purpose we consider the Su-Schrieffer-Heeger (SSH) model \cite{PhysRevLett.42.1698, PhysRevA.95.061601, PhysRevB.96.125418, PhysRevA.100.012112} for the particular choice of coupling constants. We obtain that the two mentioned above quantities are equal when the interactions between electrons are taken into account only through the dissipation rate. 

It is worth mentioning that the 1D Dirac semimetal appears as a dimensionally reduced 3D Dirac semimetal in the strong external magnetic field. Namely, in Landau level representation the lowest Landau level dominates at large enough magnetic field, and the dynamics becomes effectively one - dimensional \cite{Gusynin1999pq} (the coordinate along the direction of magnetic field is relevant). In \cite{abramchuk2024magnetoconductivity,ABRAMCHUK2026113374} it has been shown that in the presence of dissipation the chiral density and electric current are equal to each other in this limit. This, in turn, confirms the existence of a non - equilibrium version of the chiral magnetic effect {\cite{CMEZrTe5,Kharzeev2017}}. Namely, negative magnetoresistance in these materials appears to be due to the chiral magnetic effect as conjectured in \cite{CMEZrTe5,Kharzeev2017}, at least for the sufficiently large magnetic fields. 

The internal problem of the calculations presented in \cite{abramchuk2024magnetoconductivity,ABRAMCHUK2026113374} is that the effective low energy continuum theory is used instead of the tight - binding model, which describes these materials more correctly. The present paper is a step towards consideration of the negative magnetoresistance in the tight - binding models of Dirac/Weyl semimetals. We concentrate on the tight - binding model of the 1D system that, as expected, represents a result of a dimensional reduction of the 3D Dirac semimetal in the presence of strong magnetic field. As in the effective continuum theory our results confirm that the axial charge density is equal to electric current, which is a dimsnionally reduced conjecture of the chiral magnetic effect. The next step in this line of research would be the consideration of the tight - binding models of 3D Dirac semimetals with the aim at the understanding of relation between chiral disbalance and electric conductivity in the presence of electric and magnetic fields. 

In addition to the mentioned above relation to the physics of the three - dimensional materials the results obtained in the present work may have a direct meaning for the physics of the one - dimensional systems. In particular, in \cite{meier2016observation} the experimental investigation of the 1D system was reported. This system is described by the SSH model \cite{qin2023one}. Such a research may be extended in order to compare the chiral disbalance with the electric conductivity. Here the results obtained in the present paper might be relevant. Ah interesting question remains about the influence on them of disorder and interactions between electrons out of the approximation accepeted in the present paper. This, however, remains out of the scope of the present paper.   

\section{The Su-Schrieffer-Heeger (SSH) Model}
The Su-Schrieffer-Heeger (SSH) model describes a one-dimensional quantum system exhibiting topological behavior due to its alternating bond structure describing spinless electrons on a dimerized chain \cite{PhysRevLett.42.1698, PhysRevA.95.061601, PhysRevB.96.125418, PhysRevA.100.012112}. \mz{In its simplest form} the model's Hamiltonian captures electron dynamics in dimerized chains through:

\begin{equation}
H = \sum_{n} \left( J_0 c_{n,A}^\dagger c_{n,B} + J_1 c_{n+1,A}^\dagger c_{n,B} + \text{H.c.} \right),\label{H0}
\end{equation}
where $J_0$ and $J_1$ denote the intra-dimer and inter-dimer hopping integrals respectively, while $c_{n,\alpha}^\dagger$ creates an electron on sublattice $\alpha$ ($=A,B$) at unit cell $n$. This dimerized chain supports two distinct phases: a topologically non-trivial phase characterized by $|J_0| < |J_1|$ that hosts protected edge states, and a trivial phase occurring when $|J_0| > |J_1|$. The transition between these phases is marked by the closing and reopening of the energy gap, with the topological invariant being quantized to integer values of the winding number. Remarkably, domain walls between regions of different dimerization patterns give rise to soliton excitations.

The interplay between disorder and topology in these systems yields particularly intriguing phenomena. In particular, bond disorder drives the system through a topological phase transition into a topological Anderson insulator phase  \cite{PhysRevLett.42.1698, PhysRevB.108.195103,PhysRevA.100.012112, PhysRevB.96.125418}. This effect is experimentally accessible through several distinct signatures. The SSH model continues to serve as a versatile platform for exploring fundamental aspects of topological quantum matter while maintaining direct relevance to experimental realizations in diverse systems ranging from conjugated polymers to engineered photonic crystals and cold atom lattices.


In the framework of the generalized Su-Schrieffer-Heeger (SSH) model, the effects of on-site disorder on the one-dimensional (1D) chiral anomaly are rigorously studied by analyzing the system in real space. The  Hamiltonian \mz{of the generalized model} (without contribution of  disorder) is given by
\begin{equation}
    H_0 = \sum_i ( \psi_i^\dagger T_0 \psi_i + \psi_{i+1}^\dagger T_1 \psi_i + \text{H.c.} ),\label{HT}
\end{equation}
\mz{with $\psi_i = (c_{iA},c_{iB})^T $}, 
where
\begin{equation}
T_0 = \begin{pmatrix} 0 & J_0 \\ J_0 & 0 \end{pmatrix}, \quad T_1 = \begin{pmatrix} 0 & J_1 \\ J_2 & 0 \end{pmatrix},
\end{equation}
  \mz{Now  matrix $T_1$ of inter-dimer hopings is not symmetric unlike that of Eq. (\ref{H0}). Corresponsingly, parameters  $J_1$ and $J_2$ replace $J_1$ of Eq. (\ref{H0}).}  The hopping matrices are characterized by the dimerization parameter $2\Delta = J_1 - J_2$.
The dimerization parameter $\Delta$ has a strong influence on the band dispersion. For $\Delta = 0$ the dispersion (close to its minimum) is quadratic in momentum. Introducing a finite dimerization ($\Delta \neq 0$) changes this pattern. In particular, in the present paper we consider the choice of parameters corresponding to  the single gapless Dirac cone with linear dispersion.

\mz{In the folowing sections we will not be interested in effects of disorder, and will concentrate on the non - interacting version of the generalized SSH model with specific values of parameters
	\begin{equation}
		J_0 = 1, \quad J_1 = 1, \quad J_2 = 0,\label{par}
\end{equation}
{\it It is worth mentioning that in the present paper we use relativistic system of units with $\hbar = c = 1$. The values of parameters $J_i$ entering Eq. (\ref{HT}) are dimensionless. Therefore, the numerical values of Eq. (\ref{par}) are understood as expressed in a unit of energy specific for the given system. } Correspondingly, the Hamiltonian of Eq. (\ref{HT}) with the parameters of Eq. (\ref{par}) is to be multiplied by this energy unit.
Representation of this model in momentum space is considered in Appendix \ref{AppA}. One can see that actually the SSH model for the values of parameters of Eq. (\ref{par}) represents the simplest one - dimensional Wilson fermions with vanishing mass, well known in relativistic lattice field theory. We supplement the non - interacting model with the finite dissipation rate $\epsilon$, which captures somehow both effects of disorder and interactions with thermal bath of phonons. We will see using Keldysh technique that the dissipation rate in the presence of external electric field gives rise both to finite value of the accumulated axial charge density, and to the corresponding contribution to conductivity. Both these effects are related intimately to chiral anomaly.  The similar calculation has been performed for the 3+1D system in the presence of strong external magnetic field \cite{abramchuk2024magnetoconductivity}, which makes the system effectively 1+1 dimensional (including the evaluation of the dissipation rate due to disorder and interaction with phonons). However, in \cite{abramchuk2024magnetoconductivity} the effective continuum low energy effective field theory with emergent relativistic invariance was considered, while in the present paper we investigate a tight - binding model describing the real lattice systems.} 

\section{Basics of the Keldysh Technique}
The Keldysh technique is formulated to study quantum fields in nonequilibrium conditions \cite{keldysh2024diagram}, such as in the presence of external fields.  In this section we review briefly the basic notions of this technique following closely \cite{banerjee2022chiral}. Suppose the system is governed by a field Hamiltonian $\hat{H}$ and we are interested in the expectation value of an operator $\mathcal{O}[\psi,\bar{\psi}]$ constructed from fermionic fields $\psi, \bar{\psi}$ at time $t$. In the operator formalism, this average takes the form
\begin{equation}
\langle \mathcal{O} \rangle = \mathrm{tr}\!\left[ \hat{R}(t_i) e^{-i \int_{t_i}^{t} \hat{H} dt} \, \mathcal{O}[\hat{\psi},\hat{\bar{\psi}}] \, e^{-i \int_{t}^{t_f} \hat{H} dt} e^{i \int_{t_i}^{t_f} \hat{H} dt} \right],
\end{equation}
where $\hat{R}(t_i)$ is the density matrix specifying the initial state at $t_i < t$, and $t_f > t$ is a later reference time \cite{KamenevBook}.  

In the functional integral formulation, this expression is equivalently written as
\begin{equation}
\begin{split}
\langle \mathcal{O} \rangle &= \int \mathcal{D}\bar{\psi}\,\mathcal{D}\psi \mathcal{O}[\psi,\bar{\psi}] 
~~\times \\
& \exp\left\{ i \int_{\mathcal{C}} dt \int d^D x  \bar{\psi}(t,x)\,\hat{Q}\,\psi(t,x) \right\},
\end{split}
\end{equation}
where $D$ is the spatial dimension, $\mathcal{C}$ is the \textit{Keldysh contour} running forward and backward in time, and for noninteracting systems $\hat{Q}= i\partial_t - \hat{H}$. The fields defined on the forward and backward branches of the contour are independent: $\psi_\pm(t,x), \bar{\psi}_\pm(t,x)$. At the return point $t_f$, they satisfy boundary conditions $\bar{\psi}_-(t_f,x)=\bar{\psi}_+(t_f,x)$, $\psi_-(t_f,x)=\psi_+(t_f,x)$.  

The full measure of integration incorporates these conditions along with the initial distribution determined by $\hat{R}$. Explicitly,
\begin{equation}
\begin{split}
\langle \mathcal{O} \rangle &= \int \frac{\mathcal{D}\bar{\psi}_\pm \mathcal{D}\psi_\pm} {\text{Det}(1+\rho)} \mathcal{O}[\psi_+,\bar{\psi}_+] \\
&\quad \times \exp\Bigg\{ i\!\int_{t_i}^{t_f}  dt \int d^D x \big[ \bar{\psi}_-(t,x)\hat{Q}\psi_-(t,x) \\
& - \bar{\psi}_+(t,x)\hat{Q}\psi_+(t,x) \big] - \!\int d^D x \, \bar{\psi}_-(t_i,x)\rho \psi_+(t_i,x) \Bigg\},
\end{split}
\end{equation}
where $\rho$ is the single-particle density matrix characterizing the initial distribution, with occupation probabilities given by $f=\rho(1+\rho)^{-1}$.  

It is convenient to introduce the \textit{Keldysh spinor}
\begin{equation}
\Psi = \begin{pmatrix} \psi_- \\ \psi_+ \end{pmatrix},
\end{equation}
so that the expectation value becomes
\begin{equation}
\begin{split}
\langle \mathcal{O} \rangle &= \frac{1}{\text{Det}(1+\rho)} \int \mathcal{D}\bar{\Psi}\,\mathcal{D}\Psi \; \mathcal{O}[\Psi,\bar{\Psi}]\\
 & \exp\left\{ i\int_{t_i}^{t_f} dt \int d^D x \, \bar{\Psi}(t,x)\hat{Q}\Psi(t,x) \right\}.
\end{split}
\end{equation}

Here, $\hat{Q}$ is a $2\times 2$ operator matrix:

\begin{equation}
\hat{Q} = \begin{pmatrix} Q_{--} & Q_{-+} \\ Q_{+-} & Q_{++} \end{pmatrix}.
\end{equation}

For non-interacting fermions, its components are found to be
\begin{equation}
\begin{aligned}
Q_{++} &= -\Big(i\partial_t - \hat{H} - i\epsilon \tfrac{1-\rho}{1+\rho}\Big), \quad 
Q_{--} = i\partial_t - \hat{H} + i\epsilon \tfrac{1-\rho}{1+\rho}, \\
Q_{+-} &= -\tfrac{2i\epsilon}{1+\rho}, \quad 
Q_{-+} = \tfrac{2i\epsilon\,\rho}{1+\rho},
\end{aligned}
\end{equation}
with $\epsilon \to 0$ ensuring proper definition of distributions \cite{Kamenev2}.  

The \textit{Keldysh Green’s function} is then defined as
\begin{equation}
\begin{split}
G_{\alpha_1\alpha_2}(t,x|t',x') &= \int \frac{\mathcal{D}\bar{\Psi}\,\mathcal{D}\Psi}{i\text{Det}(1+\rho)} \Psi_{\alpha_1}(t,x)\bar{\Psi}_{\alpha_2}(t',x') \\
& \exp\left\{ i\int_{t_i}^{t_f} dt \int d^D x \, \bar{\Psi}(t,x)\hat{Q}\Psi(t,x) \right\},
\end{split}
\end{equation}
where $\alpha_{1,2}$ label contour branches. The Green’s function satisfies $\hat{Q}\hat{G}=1$.

A different representation of the Keldysh spinors may be introduced, which can be connected to the previously defined spinors through the following relation.
\begin{equation}
\begin{pmatrix} 
\psi_1 \\ 
\psi_2 
\end{pmatrix} = 
\frac{1}{\sqrt{2}} 
\begin{pmatrix} 
1 & 1 \\ 
1 & -1 
\end{pmatrix} 
\begin{pmatrix} 
\psi_- \\ 
\psi_+ 
\end{pmatrix},
\end{equation}
\begin{equation}
\left(\bar{\psi}_1\;\;\bar{\psi}_2\right) = 
\frac{1}{\sqrt{2}}
\left(\bar{\psi}_-\;\;\bar{\psi}_+\right)
\begin{pmatrix}
1 & 1 \\
-1 & 1
\end{pmatrix}.
\end{equation}
A convenient transformation introduces the triangular representation:
\begin{equation}
\hat{\mathbf{G}}^{(K)} = -i \left\langle
\begin{pmatrix} 
\psi_1 \\ 
\psi_2 
\end{pmatrix} 
\otimes 
\begin{pmatrix} 
\bar{\psi}_1 & \bar{\psi}_2 
\end{pmatrix} 
\right\rangle 
= 
\begin{pmatrix} 
G^{\mathrm{R}} & G^{\mathrm{K}} \\ 
0 & G^{\mathrm{A}} 
\end{pmatrix}.
\end{equation}
In our paper we will use yet another representation:

\begin{equation}
\hat{\bm{G}}^{(<)} = 
\begin{pmatrix} 
1 & 1 \\ 
0 & 1 
\end{pmatrix} 
\begin{pmatrix} 
G^{\mathrm{R}} & G^{\mathrm{K}} \\ 
0 & G^{\mathrm{A}} 
\end{pmatrix} 
\begin{pmatrix} 
1 & -1 \\ 
0 & 1 
\end{pmatrix} 
= 
\begin{pmatrix} 
G^{\mathrm{R}} & 2G^{<} \\ 
0 & G^{\mathrm{A}} 
\end{pmatrix}.
\label{eq:g_less}
\end{equation}

It is related to the Green function defined by Eq.~(11) as follows $\hat{\bm{G}}^{(<)} = U\hat{\bm{G}}V$, where 

\begin{align}
U &= \frac{1}{\sqrt{2}} 
\begin{pmatrix} 
1 & 1 \\ 
0 & 1 
\end{pmatrix} 
\begin{pmatrix} 
1 & 1 \\ 
1 & -1 
\end{pmatrix} 
= \frac{1}{\sqrt{2}} 
\begin{pmatrix} 
2 & 0 \\ 
1 & -1 
\end{pmatrix}, \\
V &= \frac{1}{\sqrt{2}} 
\begin{pmatrix} 
1 & 1 \\ 
-1 & 1 
\end{pmatrix} 
\begin{pmatrix} 
1 & -1 \\ 
0 & 1 
\end{pmatrix} 
= \frac{1}{\sqrt{2}} 
\begin{pmatrix} 
1 & 0 \\ 
-1 & 2 
\end{pmatrix}.
\end{align}
In addition, we have

\begin{equation}
\hat{\bm{Q}}^{(<)} = V^{-1}\hat{\bm{Q}}U^{-1} = 
\begin{pmatrix} 
Q^{\mathrm{R}} & 2Q^{<} \\ 
0 & Q^{\mathrm{A}} 
\end{pmatrix}.
\label{eq:q_less}
\end{equation}
Here we denote:
\begin{equation}
\begin{aligned}
Q^{\mathrm{R}} &= Q^{--} + Q^{-+}, \quad Q^{\mathrm{A}} = -Q^{-+} - Q^{++}, \\
Q^{<} &= -Q^{-+}.
\end{aligned}
\end{equation}
with the relations,
\begin{equation}
G^A = (Q^A)^{-1}, \quad G^R = (Q^R)^{-1}, \\
\quad G^< = -G^R Q^< G^A,
\end{equation}
\begin{equation}
\begin{aligned}
G^R &= (i\partial_t - \hat{H} + i\epsilon)^{-1}, \quad 
G^A = (i\partial_t - \hat{H} - i\epsilon)^{-1}, \\
G^< &= (G^A - G^R)\frac{\rho}{1+\rho}.
\end{aligned}
\end{equation}
Finally, the operator components are expressed as
\begin{equation}
\begin{aligned}
Q^< &= (Q^A - Q^R)\frac{\rho}{1+\rho} = -\frac{2i\epsilon\rho}{\rho+1}, \quad 
Q^R = i\partial_t - \hat{H} + i\epsilon, \\
Q^A &= i\partial_t - \hat{H} - i\epsilon.
\end{aligned}
\end{equation}
This compact formalism allows one to treat nonequilibrium dynamics on equal footing with equilibrium theory, reducing to Matsubara or real-time formalisms in the corresponding limits \cite{Banerjee2022snd}.

\section{Calculation of the Axial Current}

{\it We will be interested in the axial charge density and in the electric current for the particular case of vanishing chemical potential $\mu$ so that we deal with the $1D$ Fermi point.}

The calculation of the axial current begins with the Keldysh Green’s function formalism, where the expectation value of the current is written as
\begin{equation}
\langle j_5 ^\mu(x) \rangle = -i \int \frac{d^2p}{4\pi^2} \, 
\mathrm{tr}\, \sigma_3 \big[ \hat{G}_W\, \partial_{p_\mu}\hat{Q}_W \big]^{<}.
\end{equation}
Here, $\hat{G}_W$ and $\hat{Q}_W$ are Wigner - transformed \cite{Banerjee2020obs,Banerjee2022snd}  Keldysh Green function and its inverse (i.e. Dirac operator) correspondingly. The Hamiltonian is chosen as [Appendix A]
\begin{equation}
H(p_1 \mz{+} Et) = \sigma_3 \sin(p_1 \mz{+} Et) + \sigma_1 \big( 1 - \cos(p_1 \mz{+} Et) \big),\label{Hpt}
\end{equation}
with $\sigma_1$ and $\sigma_3$ being Pauli matrices. This is nothing but the one - dimensional Wilson fermions model. \mz{\it In order to express this Hamiltonian in physical units one should multiply it by the energy unit (the same as the unit of parameters in Eq. (\ref{par})). Momenta $p_1$ entering Eq. (\ref{Hpt}) are dimensionless, and are given by the physical dimensionful momenta multiplied by the lattice spacing of the given tight - binding model. Notice that we absorb the (negative) electron charge to the definition of electric field $E$, and do not include it to the definition of electric and axial currents. Therefore the obtained result for the electric conductivity is to be multiplied by $e^2$.} The Dirac operator is defined as
\begin{equation}
Q = p_0 - H(p_1 \mz{+} Et),
\end{equation}
and its derivatives with respect to momentum enter the expressions for the current. The lesser Green’s function is expressed in terms of the retarded and advanced ones as
\begin{equation}
G^< = (G^A - G^R) f(p_0) = 2\pi i \, \delta(p_0 - H)\, f(p_0),
\end{equation}
where $f(p_0)$ is the Fermi–Dirac distribution function, and the retarded and advanced Green's functions are 
\begin{equation}
G^{\mathrm{R}} = \frac{1}{p_0 - H + i\epsilon}, \quad 
G^{\mathrm{A}} = \frac{1}{p_0 - H - i\epsilon}
\end{equation}
and,

\begin{align}
Q^{<} &= (Q^{\mathrm{A}} - Q^{\mathrm{R}})f(p_0) = -2i\epsilon f(p_0).
\end{align}
where,
\begin{equation}
\begin{aligned}
Q^{\mathrm{R}} = p_0 - H + i\epsilon, \quad Q^{\mathrm{A}} = p_0 - H - i\epsilon
\end{aligned}
\end{equation}
Next, we use expansion of $\hat{\bm{G}}_W$ in powers of $F^{\mu\nu}$ and keep the terms up to the linear one,
\begin{equation}
\hat{G}_W = G^{(0)} + \frac{1}{2} F^{\mu\nu} G_{\mu\nu}^{(1)}
\label{eq:green_expansion}
\end{equation}
On the right - hand side the subscript $W$ is absent because the corresponding expressions depend on momenta and do not depend on time/coordinates.

The first-order correction to the (Wigner transformed) Green function involves terms:
\begin{equation}
\hat{G}_{\mu\nu}^{(1)} = -\hat{G} \cdot \hat{Q}_{\mu\nu}^{(1)} \cdot \hat{G} - i\left(\hat{G} \cdot \partial_{p_\mu} \hat{Q} \cdot \hat{G} \cdot \partial_{p_\nu} \hat{Q} \cdot \hat{G}\right)
\label{eq:first_order}
\end{equation}
and after tracing with $\sigma_3$, one obtains contributions that involve terms of the type
\begin{equation}
\begin{aligned}
    \mathrm{tr}\sigma_3 [\hat G_W \partial_{p_\mu} \hat Q_W]^< = \frac{i}{2} \mathrm{tr}\sigma_3 &\Big[\partial_{p_\mu} G^< \partial_{p_\nu} Q^A G^A \partial_{p_\mu} Q^A \\
    &+ \partial_{p_\mu} G^R \partial_{p_\nu} Q^< G^A \partial_{p_\mu} Q^A\\
    &+ \partial_{p_\mu} G^R \partial_{p_\nu} Q^R G^< \partial_{p_\mu} Q^A \\
    &+  \partial_{p_\mu} G^R \partial_{p_\nu} Q^R G^< \partial_{p_\mu} Q^<\Big]  F^{\mu\nu} \\
\end{aligned}
\end{equation}
where $F^{\mu\nu}$ is an external field strength.\\
Taking $\mu=0$ for the axial charge density, 
 \begin{equation}
	\begin{aligned}
		\mathrm{tr}\sigma_3 [\hat G_W \partial_{p_0} \hat Q_W]^< = \frac{i}{2} \mathrm{tr}\sigma_3 &\Big[\partial_{p_\mu} G^< \partial_{p_\nu} Q^A G^A \partial_{p_0} Q^A \\
		&+ \partial_{p_\mu} G^R \partial_{p_\nu} Q^< G^A \partial_{p_0} Q^A\\
		&+ \partial_{p_\mu} G^R \partial_{p_\nu} Q^R G^< \partial_{p_0} Q^A \\
		&+  \partial_{p_\mu} G^R \partial_{p_\nu} Q^R G^< \partial_{p_0} Q^<\Big]  F^{\mu\nu} \\
	\end{aligned}
\end{equation} 

We separate expression for the axial charge into the two parts, denoted as $I$ (Dirac sea contribution) and $II$ (Fermi surface contribution).

Dirac sea contribution $J_I$ contains integration with the Fermi - Dirac distribution.  By setting
\begin{equation}
z = p_0 - \mu,
\end{equation}
we reduce the integral over $p_0$ to the form
\begin{equation}
\begin{aligned}
\int_{-\infty}^{\infty} & \frac{dz}{e^{\beta z} + 1} 
\Bigg[
\frac{1}{(z - (E_j - \mu) - i\epsilon)^2 (z - (E_i - \mu) - i\epsilon)} \\
&- \frac{1}{(z - (E_j - \mu) + i\epsilon)^2 (z - (E_i - \mu) + i\epsilon)}
\Bigg]
\end{aligned}
\end{equation}
Residues at Matsubara poles $z = i\eta_n = 2\pi i (n+\tfrac{1}{2})/\beta$ are evaluated. In the zero-temperature limit $\beta \to \infty$, the Matsubara sum is replaced by a continuous integral. 
\begin{equation}
\begin{aligned}
\text{For } &E_i-\mu,E_j-\mu > 0 \text{ or } E_i-\mu,E_j-\mu < 0: \\
& 2 \pi \int_{-\infty}^{\infty} \frac{dz}{(z+i(E_j-\mu))^2(z+i(E_i-\mu))} = 0 
\end{aligned}
\end{equation}
\begin{equation}
\begin{aligned}
\text{For } &E_i -\mu> 0, E_j -\mu< 0:\\
&2\pi \int_{-\infty}^{\infty} \frac{dz}{(z+i(E_j-\mu))^2(z+i(E_i-\mu))} \\
& = \frac{8\pi^2 i}{(E_j-E_i)^2}
\end{aligned}
\end{equation}
\begin{equation}
\begin{aligned}
\text{For } &E_i-\mu < 0, E_j-\mu > 0:\\
&2\pi \int_{-\infty}^{\infty} \frac{dz}{(z+i(E_j-\mu))^2(z+i(E_i-\mu))} \\
& = \frac{-8\pi^2 i}{(E_j-E_i)^2}
\end{aligned}
\end{equation}

The eigenvalues and eigenvectors of the Hamiltonian are then explicitly calculated. The eigenvalues are found to be,
\begin{equation}
E_0 = -2\sin\left(\frac{P}{2}\right), \qquad 
E_1 = 2\sin\left(\frac{P}{2}\right), \qquad 
P = p_1 \mz{+} Et,
\end{equation}
with corresponding eigenvectors $|\lambda_j\rangle$ expressed in trigonometric form. From these, the matrix elements
\begin{equation}
S_{ij}(p_1) = \langle \lambda_i|\sigma_3|\lambda_j\rangle \langle \lambda_j|\partial_{p_1} H|\lambda_i\rangle
\end{equation}
are evaluated. Explicit computation shows that
\begin{equation}
S_{00} = S_{11} = \cos^2\!\left(\tfrac{P}{2}\right), \qquad 
S_{01} = S_{10} = -\sin^2\!\left(\tfrac{P}{2}\right)
\end{equation}
Plugging these into the integrals over momentum and distinguishing between the cases $E_i > \mu, E_j < \mu$ and vice versa, one finds that the terms entering $J_I$ cancel each other, so that finally  we have $J_I = 0$. 

The part $II$ contains terms proportional to $\partial_{p_0}f(p_0)$, which at zero temperature reduces to a delta function,
\begin{equation}
\partial_{p_0} f(p_0) = -\delta(p_0 - \mu)
\end{equation}
This introduces constraints on the energy levels through $\delta(p_0 - \mu)$, which ultimately reduces the double integrals into the single ones at the Fermi surface.
\begin{equation}
\begin{aligned}
J_{II}=i \int \frac{d^2 p}{4 \pi} F^{01} \, \mathrm{tr} &\Bigl[ \sigma_3 \delta (p_0 - H) \delta (p_0 - \mu) (\partial_{p_1} H)\\ &\frac{1}{p_0 - H - i \epsilon} \Bigr] 
\end{aligned}
\end{equation}
Evaluating the matrix traces with $\sigma_3$, we obtain
\begin{equation}
\begin{aligned}
J_{II}= i \int \frac{d^2 p}{4 \pi} F^{01} &\Bigl[ \braket{\lambda_i | \sigma_3 | \lambda_j} \braket{\lambda_j | \partial_{p_1} H | \lambda_i} \delta (p_0 - E_j)\\
& \delta (p_0 - \mu) \frac{1}{p_0 - E_i - i \epsilon} \Bigr]
\end{aligned}
\end{equation}
We integrate the delta-function over $p_0$
\begin{equation}
\int dp_0 \, \frac{\delta(p_0 - E_j) \delta(p_0 - \mu)}{p_0 - E_i - i\epsilon} = \frac{\delta(\mu - E_j)}{E_j - E_i - i\epsilon},   \end{equation}
which leads to,
\begin{equation}
J_{II}= \frac{i}{4\pi} F^{01} \sum_{i,j} \int dp_1 \, S_{i,j} \frac{\delta(\mu - E_j)}{E_j - E_i - i\epsilon}
\end{equation}
Considering all four choices of pairs $i,j$, we obtain:\\
For $i=0, j=0$, the integral takes the form,
\begin{equation}
J^{00}_{II}=\frac{i}{4\pi} E \int_{-\pi}^{\pi} dp_{1}  \frac{\cos^{2}\left(\frac{P}{2}\right) \delta\left(\mu - 2\sin\left(\frac{P}{2}\right)\right)}{i\epsilon}
\end{equation}
To evaluate this, we perform the change of variable $y = (p_1 \mz{+} Et)/2$, which implies $dp_1 = 2dy$. The Dirac delta function is processed using the identity:
\begin{equation}
\delta(\mu - 2\sin(y)) = \frac{1}{2 |\cos(y)|} \delta\left(y - {\rm arcsin}\,\mu/2\right)
\end{equation}
The integral over $y$ then localizes to the points where $\sin(y) = \mu/2$. In the limit of a small chemical potential ($\mu \rightarrow 0$), the evaluation yields a value of $\sqrt{1 - (\mu/2)^2}$ for the integral, which tends to 1.
\begin{equation}
J^{00}_{II}=\frac{E}{4\pi\epsilon} \sqrt{1 - \left(\frac{\mu}{2}\right)^{2}}
\approx\frac{E}{4\pi\epsilon}
\end{equation}
For $i=1, j=1$, the integral takes the form
\begin{equation}
J^{11}_{II}=\frac{i}{4\pi} E \int_{-\pi}^{\pi} dp_{1} \, \frac{\cos^{2}\left(\frac{P}{2}\right) \delta\left(\mu + 2\sin\left(\frac{P}{2}\right)\right)}{i\epsilon}
\end{equation}
In the limit of a small chemical potential, we get the same result as above,
\begin{equation}
J^{11}_{II}=\frac{E}{4\pi\epsilon} \sqrt{1 - \left(\frac{\mu}{2}\right)^{2}}
\approx\frac{E}{4\pi\epsilon}.
\end{equation}
The sum of these two terms is 
\begin{equation}
J^{11}_{II}+J^{00}_{II}=\frac{E}{2\pi\epsilon}
\end{equation}

For $i=0, j=1$, the integral takes the form,
\begin{equation}
J_{II}^{01}=\frac{i}{4\pi}E \int_{-\pi}^{\pi} \, \frac{dp_1 \sin^2\left(\frac{P}{2}\right)\delta\left(\mu - 2\sin\left(\frac{P}{2}\right)\right)}{4\sin\left(\frac{P}{2}\right) - i\epsilon}
\end{equation}
For $i=1, j=0$, the integral takes the form,
\begin{equation}
J_{II}^{10}=-\frac{i}{4\pi} E \int_{-\pi}^{\pi} \, \frac{dp_1 \sin^2\left(\frac{P}{2}\right) \delta\left(\mu + 2 \sin\left(\frac{P}{2}\right)\right)}{4 \sin\left(\frac{P}{2}\right) + i \epsilon}
\end{equation}
The sum of these terms is
\begin{equation}
J_{II}^{01}+J_{II}^{10}=\frac{-\epsilon}{32\pi} F^{01} \frac{1}{\sqrt{1-(\mu/2)^2}}
\end{equation}

We disregard the terms proportional to $\epsilon$ compared to those inverse proportional to $\epsilon$ because the dissipation rate is assumed to be small. As it was mentioned above, the contribution of the Dirac sea $J_I$ vanishes. Therefore, finally, summing all contributions  we obtain:
\begin{equation}
\langle j_5^0(x) \rangle \equiv \rho_5 = \frac{E}{2\pi\epsilon}
\end{equation}

To summarize, our calculation of the axial charge passed several steps. One starts with the Keldysh formalism, expands the Wigner transformed Green’s functions in powers of the field strength, separates contributions of Dirac sea (proportional to $f(p_0)$) and Fermi surface (proportional to  $\partial_{p_0}f(p_0)$), evaluates the resulting integrals using contour integration and Matsubara sums, diagonalizes the Hamiltonian to compute matrix elements, and finally assembles everything into the expression for the axial charge density. It appears that the contribution of Dirac sea vanishes as in \cite{abramchuk2024magnetoconductivity}. The infrared divergences are removed due to the finite dissipation rate $\epsilon$ (caused by impirities and interaction with the thermal bath of phonons). The resulting expression reveals direct proportionality between the axial charge density and the applied electric field, showing the anomaly-like structure in the calculation.

\section{Calculation of the Electric Current}

We begin from the Keldysh definition of the electric current in $1+1$ dimensions,
\begin{equation}
j \;=\; -\,i \int \frac{d^2p}{4\pi^2}\;\mathrm{tr}\,\big[\,\hat G_W\,\partial_{p_1}\hat Q_W\,\big]^<,
\end{equation}
and evaluate it to first order in the external field. Using the standard gradient (Wigner) expansion we take into account only the terms linear in the field strength,
\begin{equation}
\hat G^{(1)<}_W\;=\;\frac{i}{2}\,\big[\partial_{p_\mu}\hat G\,\partial_{p_\nu}\hat Q\,\hat G\big]^{<}F^{\mu\nu},
\end{equation}
so that

\begin{equation}
\begin{aligned}
j = -\,i\int\frac{d^2p}{4\pi^2}\;\frac{i}{2}\;\mathrm{tr}\Big[
&\partial_{p_\mu}G^{<}\,\partial_{p_\nu}Q\, G^{A}\,\partial_{p_1}Q^{A}\\
&+\partial_{p_\mu}G^{R}\,\partial_{p_\nu}Q^{<}\, G^{A}\,\partial_{p_1}Q^{A} \notag \\
&+\partial_{p_\mu}G^{R}\,\partial_{p_\nu}Q^{R}\, G^{<}\,\partial_{p_1}Q^{A}\\
&+\partial_{p_\mu}G^{R}\,\partial_{p_\nu}Q^{R}\, G^{R}\,\partial_{p_1}Q^{<}
\Big]F^{\mu\nu}
\end{aligned}
\end{equation}
Since $\partial_{p_1}Q^{<}=0$, only the first three terms contribute. We take the only nonzero field-strength component $F^{01} = - F^{10}=E$ and use $\partial_{p_1}Q^{A}=-(\partial_{p_1}H)$ together with the zeroth-order Keldysh identities.

Following an analogous calculation from the previous section, upon substitution, many terms cancel pairwise due to their symmetry. The surviving, non-canceling terms are those proportional to the energy derivative of the distribution function, $\partial_{p_0}f(p_0)$. In the zero-temperature limit, this derivative localizes the energy integration to the chemical potential $\mu$ via the relation $\partial_{p_0}f(p_0) \rightarrow -\delta(p_0 - \mu)$.

The subsequent evaluation requires taking the trace in the eigenbasis of the Hamiltonian $H(p_1 \mz{+} Et)$. The relevant matrix elements are $\langle \lambda_i | \partial_{p_1}H | \lambda_j \rangle \langle \lambda_j | \partial_{p_1}H | \lambda_i \rangle \equiv T_{i,j}$, a direct calculation yields:
\begin{align}
T_{0,0} = T_{1,1} &= \cos^2\left(\frac{P}{2}\right), \
T_{0,1} = T_{1,0} &= \sin^2\left(\frac{P}{2}\right),
\end{align}
where $P = p_1 \mz{+} Et$.

The ensuing integral over the Brillouin zone momentum $p_1 \in [-\pi, \pi]$ involves terms containing $\delta(p_0 - \mu)$ and the energy differences $(E_j - E_i \pm i\epsilon)$. The calculation reveals that the contributions from the off-diagonal matrix elements $T_{0,1}$ and $T_{1,0}$ are proportional to the infinitesimal regulator $\epsilon$ in the numerator. Consequently, in the limit $\epsilon \rightarrow 0$ (taken at the end of the calculation), these terms vanish identically, leaving only the contributions from the diagonal elements $T_{0,0}$ and $T_{1,1}$.

The two remaining non-vanishing integrals are identical and take the form:
\begin{equation}
\frac{i}{4\pi} E \int_{-\pi}^{\pi} dp_1 , \frac{\cos^2\left(\frac{P}{2}\right) \delta(\mu - 2\sin(\frac{P}{2}))}{i\epsilon}
\end{equation}
Similarly, summing the two identical contributions yields the final result for the electric current:
\begin{equation}
\langle j \rangle = \frac{E}{2\pi\epsilon}
\end{equation}
This result is identical to that obtained for the axial charge density $\rho_5$ derived previously. Therefore, we conclude that within this model and to first order in the field strength, the electrically induced current and the axial charge density are equal at $\mu = 0$, and their difference vanishes:
\begin{equation}
j - \rho_5 = 0
\end{equation}

\section{Chiral Anomaly}

\mz{In this section, we calculate the chiral anomaly in the considered model in two different ways. First of all, we use Kubo formulas, i.e. calculate the response of axial current (and its derivative) to external electric field within the equilibrium Matsubara technique in its form based on Wigner - Weyl calculus. In this calculation we adopt the gauge in which the fields do not depend on time. 	Next, we obtain the same result in the other gauge, in which vector potential depends on time. Then we  derive the chiral anomaly  using the Keldysh nonequilibrium Green’s function formalism.} 

\subsection{Calculation via Kubo formula}

Here we follow the approach developed in \cite{zubkov2017topology}, in its simplified form adopted for the $1+1$ D models.
We begin by considering an inhomogeneous lattice system \mz{at zero temperature} coupled to electric potential depending on coordinates but independent of time. The chiral current density is expressed as:

\begin{equation}
\begin{aligned}
j^\rho_{5}(x) = -\int_{\mathcal{M}} \frac{d^2 p}{|\mathcal{M}|}  \mathrm{tr} &\bigl( \gamma^5 G_W(x, p) \\
&\partial_{p_\rho} Q_W(x, p - A(x)) \bigr),
\end{aligned}
\end{equation}
\mz{where $ |\mathcal{M}| = (2\pi)^2$ is volume of momentum space}, \( G_W \) and \( Q_W \) are the Wigner-transformed \mz{Matsubara} Green’s function and Weyl symbol of Dirac operator, respectively, and \( \gamma^5=\sigma^3 \) is $1+1$ D chirality operator. To extract the anomalous divergence, we consider the spatially averaged expectation value of the divergence of the current:
\begin{equation}
\overline{\langle \partial j_5 \rangle} = \frac{1}{\mathbf{L}} \int d x  \partial_\mu j_5^\mu(x),
\end{equation}

\mz{where  \( \mathbf{L} \) is the length of the $1D$ system. We obtain:
\begin{equation}
\begin{aligned}
\overline{\langle \partial j_5 \rangle} = -\frac{1}{ \mathbf{L}} \int d x \int_{\mathcal{M}} \frac{d^2 p}{|\mathcal{M}|}  \mathrm{tr} &\bigl( \gamma^5 \partial_{p_2} \left( G \partial_{p_\rho} Q \right) \bigr) \\
&\partial_\rho A_{2}(x) 
\end{aligned}
\end{equation}
Here $A_2$ is the temporal component of vector potential after Wick rotation. We omit subscript $W$ because both Dirac operator and the Green function loose coordinate dependence, when linear response of the axial current to electric field is considered.  Integral over $p_2$ is nonzero because of the pole of expression standing inside the integral. Using regularization of the theory by finite temperature we arrive at
\begin{equation}
\begin{aligned}
&\overline{\langle \partial j_5 \rangle}  = \frac{\cal N}{\pi} E,\\ & {\cal N } = \frac{1}{4\pi i }  \int_{\cal C}  \mathrm{tr} \bigl( \gamma^5 G d Q \bigr) 
\end{aligned}\label{topinv}
\end{equation}
Here $E$ is external electric field (assumed to be constant), contour $\cal C$ embraces the position of Fermi point in momentum space. One can see that the topological invariant $\cal N$ is equal to $1$ for the considered model. 
This expression captures the chiral anomaly: the divergence of the chiral current is proportional to the electric field. Below we will see that it  arises from the spectral flow across the Fermi point.}

\subsection{Calculation using Keldysh technique}

Here we consider the gauge \( A_{0} = 0 \), \( A_{1} = -Et \), where \( E \) is  constant electric field. The Hamiltonian takes the form:
\begin{equation}
H(p_1 \mz{+} Et) = \sigma_3 \sin(p_1 \mz{+} Et) + \sigma_1 (1 - \cos(p_1 \mz{+} Et))
\end{equation}
and the axial current is given by:
\begin{equation}
\langle j_5^{\mu}(x) \rangle = -i \int \frac{d^{2} p}{4\pi^{2}}  \mathrm{tr}  \sigma_3 G^{<} \partial_{p_{\mu}} Q
\end{equation}
with \( Q = p_0 - H \). Using the expressions for the retarded, advanced, and lesser Green’s functions and their counterparts for \( Q \) and \( \partial_{p_0} Q = 1\), we simplify the current expectation value. The averaged divergence becomes:
\begin{equation}
\overline{\langle \partial j_5 \rangle} = -{i }  \int \frac{d^{2} p}{4\pi^{2}}  \partial_{t}  \mathrm{tr}  \sigma_3 G^{<}
\end{equation}

To evaluate this, we employ an adiabatic approximation valid when $Et$ is sufficiently small. The eigenstates of \( H \) in momentum ($p=p_1$) representation are:
\begin{equation}
\begin{aligned}
&|+, P \rangle = (\cos(P/4), \sin(P/4))^T e^{ i\phi_+(t,E,p)}, \\
&E_+ = 2 \sin(P/2), \\
&|-, P \rangle = (\sin(P/4), -\cos(P/4))^T e^{ i\phi_-(t,E,p)}, \\
&E_- = -2 \sin(P/2)
\end{aligned}
\end{equation}
where $\phi_\pm$ is the real - valued phase that governs the evolution in time in adiabatic approximation. These states are simultaneously the eigenstates of momentum with eigenvalues \( P = p \mz{+} Et \) and (approximately) the eigenstates of the time depending Hamiltonian corresponding to the time depending energy values $E_\pm$. $p$ is the value of momentum at $t = 0$. Notice that this is $P$ that has the meaning of particle momentum at $t\ne 0$ rather than $p$. The lesser Green’s function is constructed from these states, incorporating a step function that tracks the time-dependent occupation at zero temperature:
\begin{equation}
G^{<}(p) = 2\pi i \sum_{\pm} |\pm, P \rangle \langle \pm, P|  \delta(p_0 - E_{\pm}(P))  \theta(-E_{\pm}(P - Et)).
\end{equation}
Substituting into the expression for \( \overline{\langle \partial j \rangle} \), we find:
\begin{equation}
\begin{aligned}
\overline{\langle \partial j_5 \rangle} =  \partial_t  &\int_{-\pi}^{\pi} \frac{dp}{2\pi}  \mathrm{tr}  \sigma_3 \\
&\sum_{\pm} |\pm, P \rangle \langle \pm, P|  \theta(-E_{\pm}(P- Et))
\end{aligned}
\end{equation}
Evaluating the trace and the integral over \( P \), we arrive at:
\begin{equation}
\overline{\langle \partial j_5 \rangle} = \frac{E}{\pi} \cos(Et/2)\approx \frac{E}{\pi}
\end{equation}
The very approximation we used is valid for $Et \ll 1$. Therefore, we take the limit \( t \to 0 \), this reduces to \( E/\pi \), which coincides with the above obtained expression. This result illustrates how the electric field drives a spectral flow between the two chiral branches, leading to a non-conservation of the chiral charge. 

\begin{figure}[h]
    \centering
    \includegraphics[width=0.8\linewidth]{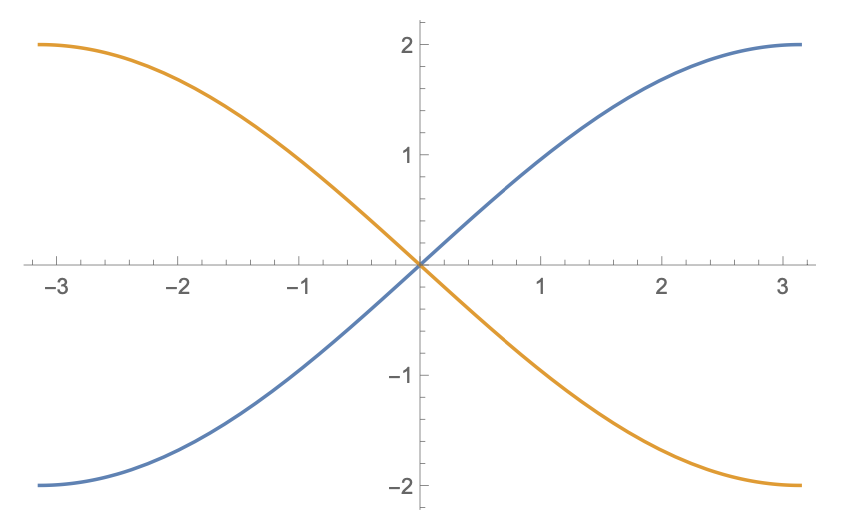}
    \caption{$E_{1,2}$ as a function of $p$}
    \label{figure1}
\end{figure}

The energy eigenvalues $E_{1,2}(p) = \pm 2\sin(p/2)$, depicted in Fig.\ref{figure1}, form a sinusoidal, periodic structure over the Brillouin zone. The application of a constant electric field $E$ induces a linear drift in crystal momentum, $P = p \mz{+} Et$, causing the quantum states to adiabatically traverse this band structure. This spectral flow results in a continuous transfer of states from the negative-energy sea to the positive-energy continuum, which manifests as the non-conservation of chiral charge and provides the microscopic mechanism for the chiral anomaly $\overline{\langle \partial j \rangle} = E/\pi$ in this lattice model.

This derivation underscores the power of the Keldysh formalism in capturing non-equilibrium quantum anomalies and provides a robust framework for studying chiral transport in driven quantum systems. The result not only confirms the field-theoretic expectation but also offers insight into the microscopic mechanisms behind anomaly-induced transport in lattice models.

\section{Conclusions}

\mz{To conclude in the present paper we analyze the generalized SSH model neglecting interactions between the fermions and disorder. We take into account, however, the effect of both disorder and interactions implementing the finite (but small) dissipation rate.}  
We have derived the dynamical response of the chiral charge and the electrical current in the Su--Schrieffer--Heeger model to external electric field, using the Keldysh non-equilibrium formalism. The dissipation rate $\epsilon$ is inverse to the relaxation time $\tau = 1/(2\epsilon)$. \mz{We chose specific values of parameters, which provide appearance of the Dirac point, and considered the case when Fermi energy is tuned to the position of the Dirac point (i.e. when $\mu \to 0$).} 

In the absence of disorder and dissipation, the rate of chiral charge production would be given by
\begin{equation}
\overline{\langle \partial j \rangle} = \partial_t \rho_5 = \frac{E}{\pi}
\end{equation}
 in the limit $t \to 0$.
This expression proposed the linear growth of the chiral imbalance at least at small values of $t$:
\begin{equation}
\rho_5 = \frac{E t}{\pi}
\end{equation}

When dissipation is introduced via a finite relaxation rate $\epsilon$, the chiral charge saturates to a steady-state value proportional to the relaxation time:
\begin{equation}
\rho_5 = \frac{E}{\pi} \tau, \quad \text{where} \quad \tau = \frac{1}{2\epsilon}
\end{equation}

Remarkably, the electric current $j$ exhibits the same functional dependence on electric field and relaxation time. We find:
\begin{equation}
j = \rho_5 = \frac{E}{\pi} \tau,
\end{equation}
which implies a direct proportionality between the current and the chiral imbalance. Consequently, the electrical conductivity is given by
\begin{equation}
\sigma = \frac{j}{E} = \frac{\tau}{\pi}\label{condfin}
\end{equation}

This result demonstrates that the anomaly-induced contribution to the conductivity is explicitly governed by the relaxation time $\tau$, highlighting the interplay between topological spectral flow and dissipative processes in the non-equilibrium steady state.

One of the interesting questions is relation to topology of the coefficient standing in expression for the conductivity of Eq. (\ref{condfin}). In the considered model the coefficient at the ratio $\frac{\tau}{\pi}$ is simply equal to unity. For the  models with several species of $1+1$D Dirac fermions with linear dispersion this number will be equal to the number of these species. One may expect that in the general case of an arbitrary one - dimensional Dirac semimetal such a coefficient is equal to a topological invariant of Eq. (\ref{topinv}). Its value may differ from the number of the species of $1+1$ D Dirac spinors if the dispersion is more complicated \cite{Volovik2003a}. The derivation of this relation may be performed using Wigner - Weyl calculus (see, for example, \cite{chernodub2017scale,zhang2020influence}), but this is out of the scope of the present paper. Notice that the similar expression for the chiral separation effect in the three - dimensional systems  \cite{zubkov2023effect} contains a topological invariant composed of the Green  function. According to the general approach of \cite{Volovik2003a} such topological invariants provide connection between solid state systems and high energy physics  \cite{zubkov2018momentum,zubkov2012momentum,volovik2017standard}. This analogy may be extended also to the fermionic superfluids \cite{volovik2013nambu}. 

\section{Appendix}
\subsection{Hamiltonian of generalized SSH model in momentum space}
\label{AppA}
We begin with the general tight-binding Hamiltonian for a one-dimensional chain with two sites per unit cell:
\begin{equation}
H_0 = \sum_i \left( \psi_i^\dagger T_0 \psi_i + \psi_{i+1}^\dagger T_1 \psi_i + \text{H.c.} \right),
\end{equation}
where $\psi = \begin{pmatrix} a_{i} \ b_{i} \end{pmatrix}^T$, and the hopping matrices are:
\begin{equation}
T_0 = \begin{pmatrix} 0 & J_0 \\ J_0 & 0 \end{pmatrix}, \quad T_1 = \begin{pmatrix} 0 & J_1 \\ J_2 & 0 \end{pmatrix}
\end{equation}

\noindent Performing a Fourier transformation using:
\begin{equation}
a_i = \frac{1}{\sqrt{N}} \sum_k e^{ikx_i} a_k, \quad b_i = \frac{1}{\sqrt{N}} \sum_k e^{ikx_i} b_k,
\end{equation}
we obtain the momentum-space Hamiltonian:
\begin{equation}
H_0 = \sum_k \begin{pmatrix} a_k^\dagger & b_k^\dagger \end{pmatrix} \mathcal{H}(k) \begin{pmatrix} a_k \\ b_k \end{pmatrix},
\end{equation}
where the Bloch Hamiltonian is:
\begin{equation}
\mathcal{H}(k) = \begin{pmatrix} 0 & J_0 + J_1 e^{-ik} + J_2 e^{ik} \\ J_0 + J_2 e^{-ik} + J_1 e^{ik} & 0 \end{pmatrix}
\end{equation}

\noindent This can be expressed in Pauli matrix form as:
\begin{equation}
\mathcal{H}(k) = h_x(k) \sigma_1 + h_y(k) \sigma_2,
\end{equation}
with components:
\begin{equation}
\begin{aligned}
&h_x(k) = J_0 + (J_1 + J_2)\cos(k), \\
&h_y(k) = (J_1 - J_2)\sin(k)
\end{aligned}
\end{equation}

\noindent To obtain a minimal model for \mz{ 1D topological semimetal}, we select specific parameters:
\begin{equation}
J_0 = 1, \quad J_1 = 1, \quad J_2 = 0,
\end{equation}
which yields:
\begin{equation}
\begin{aligned}
h_x(k) &= 1 + \cos(k), \\
h_y(k) &= \sin(k),
\end{aligned}
\end{equation}
resulting in:
\begin{equation}
\mathcal{H}(k) = (1 + \cos(k)) \sigma_1 + \sin(k) \sigma_2
\end{equation}

\noindent We perform a basis transformation using the unitary operator $U = e^{i(\pi/4)\sigma_2} = \frac{1}{\sqrt{2}} \begin{pmatrix} 1 & 1 \\ -1 & 1 \end{pmatrix}$, under which:
\begin{equation}
\begin{aligned}
U \sigma_1 U^\dagger &= \sigma_1, \\
U \sigma_2 U^\dagger &= \sigma_3
\end{aligned}
\end{equation}
Applying this transformation:
\begin{equation}
\begin{aligned}
H(k) &= U \mathcal{H}(k) U^\dagger \\
&= (1 + \cos(k)) (U \sigma_1 U^\dagger) + \sin(k) (U \sigma_2 U^\dagger) \\
&= (1 + \cos(k)) \sigma_1 + \sin(k) \sigma_3
\end{aligned}
\end{equation}
\noindent Applying a momentum shift $k \rightarrow k + \pi$ and sign reversal:
\begin{equation}
\begin{aligned}
H(k+\pi) &= (1 + \cos(k+\pi)) \sigma_1 + \sin(k+\pi) \sigma_3 \\
&= (1 - \cos(k)) \sigma_1 - \sin(k) \sigma_3, \\
-H(k+\pi) &= \sin(k) \sigma_3 + (1 - \cos(k)) \sigma_1.
\end{aligned}
\end{equation}
Thus, we define our final Hamiltonian as:
\begin{equation}
H_{\text{final}}(k) = \sin(k) \sigma_3 + (1 - \cos(k)) \sigma_1
\end{equation}
This is the Well - known in lattice field theory form of the Euclidean one - dimensional Wilson fermion's action. Specific choice of parameters gives rise to Dirac point at $k = 0$.  

\noindent Finally, we incorporate the electric field $E$ via the Peierls substitution $k \rightarrow k \mz{+} Et$:
\begin{equation}
H(k - Et) = \sigma_3 \sin(k \mz{+} Et) + \sigma_1 (1 - \cos(k \mz{+} Et))
\end{equation}
Recognizing $k$ as the momentum $p_1$, we obtain the desired result:
\begin{equation}
H(p_1 \mz{+} Et) = \sigma_3 \sin(p_1 \mz{+} Et) + \sigma_1 \big( 1 - \cos(p_1 \mz{+} Et) \big)
\end{equation}

\bibliographystyle{apsrev4-2}
\bibliography{refs,CMEwKeldysh_ext,CSE_MZ,references,biblio_corrected,wigner3,cross-ref}

\begin{thebibliography}{76}%
\makeatletter
\providecommand \@ifxundefined [1]{%
 \@ifx{#1\undefined}
}%
\providecommand \@ifnum [1]{%
 \ifnum #1\expandafter \@firstoftwo
 \else \expandafter \@secondoftwo
 \fi
}%
\providecommand \@ifx [1]{%
 \ifx #1\expandafter \@firstoftwo
 \else \expandafter \@secondoftwo
 \fi
}%
\providecommand \natexlab [1]{#1}%
\providecommand \enquote  [1]{``#1''}%
\providecommand \bibnamefont  [1]{#1}%
\providecommand \bibfnamefont [1]{#1}%
\providecommand \citenamefont [1]{#1}%
\providecommand \href@noop [0]{\@secondoftwo}%
\providecommand \href [0]{\begingroup \@sanitize@url \@href}%
\providecommand \@href[1]{\@@startlink{#1}\@@href}%
\providecommand \@@href[1]{\endgroup#1\@@endlink}%
\providecommand \@sanitize@url [0]{\catcode `\\12\catcode `\$12\catcode
  `\&12\catcode `\#12\catcode `\^12\catcode `\_12\catcode `\%12\relax}%
\providecommand \@@startlink[1]{}%
\providecommand \@@endlink[0]{}%
\providecommand \url  [0]{\begingroup\@sanitize@url \@url }%
\providecommand \@url [1]{\endgroup\@href {#1}{\urlprefix }}%
\providecommand \urlprefix  [0]{URL }%
\providecommand \Eprint [0]{\href }%
\providecommand \doibase [0]{https://doi.org/}%
\providecommand \selectlanguage [0]{\@gobble}%
\providecommand \bibinfo  [0]{\@secondoftwo}%
\providecommand \bibfield  [0]{\@secondoftwo}%
\providecommand \translation [1]{[#1]}%
\providecommand \BibitemOpen [0]{}%
\providecommand \bibitemStop [0]{}%
\providecommand \bibitemNoStop [0]{.\EOS\space}%
\providecommand \EOS [0]{\spacefactor3000\relax}%
\providecommand \BibitemShut  [1]{\csname bibitem#1\endcsname}%
\let\auto@bib@innerbib\@empty
\bibitem [{\citenamefont {Onoda}\ \emph {et~al.}(6 07)\citenamefont {Onoda},
  \citenamefont {Sugimoto},\ and\ \citenamefont {Nagaosa}}]{Onoda2006}%
  \BibitemOpen
  \bibfield  {author} {\bibinfo {author} {\bibfnamefont {S.}~\bibnamefont
  {Onoda}}, \bibinfo {author} {\bibfnamefont {N.}~\bibnamefont {Sugimoto}},\
  and\ \bibinfo {author} {\bibfnamefont {N.}~\bibnamefont {Nagaosa}},\ }\href
  {https://doi.org/10.1143/PTP.116.61} {\bibfield  {journal} {\bibinfo
  {journal} {Progress of Theoretical Physics}\ }\textbf {\bibinfo {volume}
  {116}},\ \bibinfo {pages} {61} (\bibinfo {year} {2006-07})}\BibitemShut
  {NoStop}%
\bibitem [{\citenamefont {Su}\ \emph {et~al.}(1979)\citenamefont {Su},
  \citenamefont {Schrieffer},\ and\ \citenamefont
  {Heeger}}]{PhysRevLett.42.1698}%
  \BibitemOpen
  \bibfield  {author} {\bibinfo {author} {\bibfnamefont {W.~P.}\ \bibnamefont
  {Su}}, \bibinfo {author} {\bibfnamefont {J.~R.}\ \bibnamefont {Schrieffer}},\
  and\ \bibinfo {author} {\bibfnamefont {A.~J.}\ \bibnamefont {Heeger}},\
  }\href {https://doi.org/10.1103/PhysRevLett.42.1698} {\bibfield  {journal}
  {\bibinfo  {journal} {Phys. Rev. Lett.}\ }\textbf {\bibinfo {volume} {42}},\
  \bibinfo {pages} {1698} (\bibinfo {year} {1979})}\BibitemShut {NoStop}%
\bibitem [{\citenamefont {Qin}\ \emph {et~al.}(2023{\natexlab{a}})\citenamefont
  {Qin}, \citenamefont {Xu}, \citenamefont {Ning},\ and\ \citenamefont
  {Wang}}]{PhysRevB.108.195103}%
  \BibitemOpen
  \bibfield  {author} {\bibinfo {author} {\bibfnamefont {Z.}~\bibnamefont
  {Qin}}, \bibinfo {author} {\bibfnamefont {D.-H.}\ \bibnamefont {Xu}},
  \bibinfo {author} {\bibfnamefont {Z.}~\bibnamefont {Ning}},\ and\ \bibinfo
  {author} {\bibfnamefont {R.}~\bibnamefont {Wang}},\ }\href
  {https://doi.org/10.1103/PhysRevB.108.195103} {\bibfield  {journal} {\bibinfo
   {journal} {Phys. Rev. B}\ }\textbf {\bibinfo {volume} {108}},\ \bibinfo
  {pages} {195103} (\bibinfo {year} {2023}{\natexlab{a}})}\BibitemShut
  {NoStop}%
\bibitem [{\citenamefont {Du}\ \emph {et~al.}(2019)\citenamefont {Du},
  \citenamefont {Wu}, \citenamefont {Artoni},\ and\ \citenamefont
  {La~Rocca}}]{PhysRevA.100.012112}%
  \BibitemOpen
  \bibfield  {author} {\bibinfo {author} {\bibfnamefont {L.}~\bibnamefont
  {Du}}, \bibinfo {author} {\bibfnamefont {J.-H.}\ \bibnamefont {Wu}}, \bibinfo
  {author} {\bibfnamefont {M.}~\bibnamefont {Artoni}},\ and\ \bibinfo {author}
  {\bibfnamefont {G.~C.}\ \bibnamefont {La~Rocca}},\ }\href
  {https://doi.org/10.1103/PhysRevA.100.012112} {\bibfield  {journal} {\bibinfo
   {journal} {Phys. Rev. A}\ }\textbf {\bibinfo {volume} {100}},\ \bibinfo
  {pages} {012112} (\bibinfo {year} {2019})}\BibitemShut {NoStop}%
\bibitem [{\citenamefont {Li}\ \emph {et~al.}(2017)\citenamefont {Li},
  \citenamefont {Lin}, \citenamefont {Zhang},\ and\ \citenamefont
  {Song}}]{PhysRevB.96.125418}%
  \BibitemOpen
  \bibfield  {author} {\bibinfo {author} {\bibfnamefont {C.}~\bibnamefont
  {Li}}, \bibinfo {author} {\bibfnamefont {S.}~\bibnamefont {Lin}}, \bibinfo
  {author} {\bibfnamefont {G.}~\bibnamefont {Zhang}},\ and\ \bibinfo {author}
  {\bibfnamefont {Z.}~\bibnamefont {Song}},\ }\href
  {https://doi.org/10.1103/PhysRevB.96.125418} {\bibfield  {journal} {\bibinfo
  {journal} {Phys. Rev. B}\ }\textbf {\bibinfo {volume} {96}},\ \bibinfo
  {pages} {125418} (\bibinfo {year} {2017})}\BibitemShut {NoStop}%
\bibitem [{\citenamefont {Keldysh}(1964)}]{Keldysh64}%
  \BibitemOpen
  \bibfield  {author} {\bibinfo {author} {\bibfnamefont {L.~V.}\ \bibnamefont
  {Keldysh}},\ }\href@noop {} {\bibfield  {journal} {\bibinfo  {journal} {Zh.
  Eksp. Teor. Fiz. 47 (1964), 1515 [Sov. Phys. JETP 20 (1965), 1018].}\
  }\textbf {\bibinfo {volume} {47}},\ \bibinfo {pages} {1515} (\bibinfo {year}
  {1964})}\BibitemShut {NoStop}%
\bibitem [{\citenamefont {Keldysh}(2024)}]{keldysh2024diagram}%
  \BibitemOpen
  \bibfield  {author} {\bibinfo {author} {\bibfnamefont {L.~V.}\ \bibnamefont
  {Keldysh}},\ }in\ \href@noop {} {\emph {\bibinfo {booktitle} {Selected Papers
  of Leonid V Keldysh}}}\ (\bibinfo  {publisher} {World Scientific},\ \bibinfo
  {year} {2024})\ pp.\ \bibinfo {pages} {47--55}\BibitemShut {NoStop}%
\bibitem [{\citenamefont {Schwinger}(1961)}]{Schwinger61}%
  \BibitemOpen
  \bibfield  {author} {\bibinfo {author} {\bibfnamefont {J.}~\bibnamefont
  {Schwinger}},\ }\href@noop {} {\bibfield  {journal} {\bibinfo  {journal} {J.
  Math. Phys. 2 (1961), 407.}\ }\textbf {\bibinfo {volume} {2}},\ \bibinfo
  {pages} {407} (\bibinfo {year} {1961})}\BibitemShut {NoStop}%
\bibitem [{\citenamefont {Banerjee}\ \emph {et~al.}(2021)\citenamefont
  {Banerjee}, \citenamefont {Fialkovsky}, \citenamefont {Lewkowicz},
  \citenamefont {Zhang},\ and\ \citenamefont {Zubkov}}]{Banerjee2020obs}%
  \BibitemOpen
  \bibfield  {author} {\bibinfo {author} {\bibfnamefont {C.}~\bibnamefont
  {Banerjee}}, \bibinfo {author} {\bibfnamefont {I.~V.}\ \bibnamefont
  {Fialkovsky}}, \bibinfo {author} {\bibfnamefont {M.}~\bibnamefont
  {Lewkowicz}}, \bibinfo {author} {\bibfnamefont {C.~X.}\ \bibnamefont
  {Zhang}},\ and\ \bibinfo {author} {\bibfnamefont {M.~A.}\ \bibnamefont
  {Zubkov}},\ }\href {https://doi.org/10.1007/s10825-021-01775-8} {\bibfield
  {journal} {\bibinfo  {journal} {Electronics}\ }\textbf {\bibinfo {volume}
  {20}},\ \bibinfo {pages} {2255} (\bibinfo {year} {2021})},\ \Eprint
  {https://arxiv.org/abs/2009.10704} {arXiv:2009.10704 [cond-mat.mes-hall]}
  \BibitemShut {NoStop}%
\bibitem [{\citenamefont {Banerjee}\ \emph
  {et~al.}(2022{\natexlab{a}})\citenamefont {Banerjee}, \citenamefont
  {Lewkowicz},\ and\ \citenamefont {Zubkov}}]{Banerjee2022snd}%
  \BibitemOpen
  \bibfield  {author} {\bibinfo {author} {\bibfnamefont {C.}~\bibnamefont
  {Banerjee}}, \bibinfo {author} {\bibfnamefont {M.}~\bibnamefont
  {Lewkowicz}},\ and\ \bibinfo {author} {\bibfnamefont {M.~A.}\ \bibnamefont
  {Zubkov}},\ }\href {https://doi.org/10.1103/PhysRevD.106.074508} {\bibfield
  {journal} {\bibinfo  {journal} {Phys. Rev. D}\ }\textbf {\bibinfo {volume}
  {106}},\ \bibinfo {pages} {074508} (\bibinfo {year} {2022}{\natexlab{a}})},\
  \Eprint {https://arxiv.org/abs/2206.11819} {arXiv:2206.11819 [hep-ph]}
  \BibitemShut {NoStop}%
\bibitem [{\citenamefont {Arseev}(2015)}]{Arseev2015}%
  \BibitemOpen
  \bibfield  {author} {\bibinfo {author} {\bibfnamefont {P.~I.}\ \bibnamefont
  {Arseev}},\ }\href {https://doi.org/10.3367/UFNe.0185.201512b.1271}
  {\bibfield  {journal} {\bibinfo  {journal} {Phys. Usp.}\ }\textbf {\bibinfo
  {volume} {58}},\ \bibinfo {pages} {1159} (\bibinfo {year}
  {2015})}\BibitemShut {NoStop}%
\bibitem [{\citenamefont {Kamenev}(2005)}]{Kamenev2005course}%
  \BibitemOpen
  \bibfield  {author} {\bibinfo {author} {\bibfnamefont {A.}~\bibnamefont
  {Kamenev}},\ }in\ \href@noop {} {\emph {\bibinfo {booktitle} {Les
  Houches}}},\ Vol.~\bibinfo {volume} {81}\ (\bibinfo  {publisher} {Elsevier},\
  \bibinfo {year} {2005})\ pp.\ \bibinfo {pages} {177--246},\ \Eprint
  {https://arxiv.org/abs/cond-mat/0412296} {arXiv:cond-mat/0412296
  [cond-mat.dis-nn]} \BibitemShut {NoStop}%
\bibitem [{\citenamefont {Kamenev}(2023)}]{KamenevBook}%
  \BibitemOpen
  \bibfield  {author} {\bibinfo {author} {\bibfnamefont {A.}~\bibnamefont
  {Kamenev}},\ }\href@noop {} {\emph {\bibinfo {title} {Field theory of
  non-equilibrium systems}}}\ (\bibinfo  {publisher} {Cambridge University
  Press},\ \bibinfo {year} {2023})\BibitemShut {NoStop}%
\bibitem [{\citenamefont {Kadanoff}\ and\ \citenamefont {Baym}()}]{KB62}%
  \BibitemOpen
  \bibfield  {author} {\bibinfo {author} {\bibfnamefont {L.~P.}\ \bibnamefont
  {Kadanoff}}\ and\ \bibinfo {author} {\bibfnamefont {G.}~\bibnamefont
  {Baym}},\ }\href@noop {} {\bibinfo {title} {{Quantum Statistical
  Mechanics}}},\ \bibinfo {note} {(Benjamin, New York, 1962).}\BibitemShut
  {Stop}%
\bibitem [{\citenamefont {Baym}(1962)}]{Baym62}%
  \BibitemOpen
  \bibfield  {author} {\bibinfo {author} {\bibfnamefont {G.}~\bibnamefont
  {Baym}},\ }\href@noop {} {\bibfield  {journal} {\bibinfo  {journal} {Phys.
  Rev. 127 (1962), 1391.}\ }\textbf {\bibinfo {volume} {127}},\ \bibinfo
  {pages} {1391} (\bibinfo {year} {1962})}\BibitemShut {NoStop}%
\bibitem [{\citenamefont {Matsubara}(1955)}]{Matsubara55}%
  \BibitemOpen
  \bibfield  {author} {\bibinfo {author} {\bibfnamefont {T.}~\bibnamefont
  {Matsubara}},\ }\href@noop {} {\bibfield  {journal} {\bibinfo  {journal}
  {Prog. Theor. Phys. 14 (1955), 351.}\ }\textbf {\bibinfo {volume} {14}},\
  \bibinfo {pages} {351} (\bibinfo {year} {1955})}\BibitemShut {NoStop}%
\bibitem [{\citenamefont {Bloch}\ and\ \citenamefont
  {de~Donimicis}(1959)}]{BdD59}%
  \BibitemOpen
  \bibfield  {author} {\bibinfo {author} {\bibfnamefont {C.}~\bibnamefont
  {Bloch}}\ and\ \bibinfo {author} {\bibfnamefont {C.}~\bibnamefont
  {de~Donimicis}},\ }\href@noop {} {\bibfield  {journal} {\bibinfo  {journal}
  {Nucl. Phys. 10 (1959), 509.}\ }\textbf {\bibinfo {volume} {10}},\ \bibinfo
  {pages} {509} (\bibinfo {year} {1959})}\BibitemShut {NoStop}%
\bibitem [{\citenamefont {Gaudin}(1960)}]{Gaudin60}%
  \BibitemOpen
  \bibfield  {author} {\bibinfo {author} {\bibfnamefont {M.}~\bibnamefont
  {Gaudin}},\ }\href@noop {} {\bibfield  {journal} {\bibinfo  {journal} {Nucl.
  Phys. 15 (1960), 89.}\ }\textbf {\bibinfo {volume} {15}},\ \bibinfo {pages}
  {89} (\bibinfo {year} {1960})}\BibitemShut {NoStop}%
\bibitem [{\citenamefont {A.~A.~Abrikosov}\ and\ \citenamefont
  {Dzyaloshinski}()}]{AGD63}%
  \BibitemOpen
  \bibfield  {author} {\bibinfo {author} {\bibfnamefont {L.~P.~G.}\
  \bibnamefont {A.~A.~Abrikosov}}\ and\ \bibinfo {author} {\bibfnamefont
  {I.~E.}\ \bibnamefont {Dzyaloshinski}},\ }\href@noop {} {\bibinfo {title}
  {{Methods of Quantum Field Theory in Statistical Physics}}},\ \bibinfo {note}
  {(Prentice-Hall, Englewood Cliffs, N.J., 1963).}\BibitemShut {Stop}%
\bibitem [{\citenamefont {Langreth}()}]{Langreth76}%
  \BibitemOpen
  \bibfield  {author} {\bibinfo {author} {\bibfnamefont {D.~C.}\ \bibnamefont
  {Langreth}},\ }\href@noop {} {\bibinfo {title} {{in Linear and Nonlinear
  Electron Transport in Solids}}},\ \bibinfo {note} {eds. J. T. Devreese and V.
  E. van Doren (Plenum Press, New York, 1976), p. 3.}\BibitemShut {Stop}%
\bibitem [{\citenamefont {Danielewicz}(1984)}]{Danielewicz84}%
  \BibitemOpen
  \bibfield  {author} {\bibinfo {author} {\bibfnamefont {P.}~\bibnamefont
  {Danielewicz}},\ }\href@noop {} {\bibfield  {journal} {\bibinfo  {journal}
  {Ann. of Phys. 152 (1984), 239.}\ }\textbf {\bibinfo {volume} {152}},\
  \bibinfo {pages} {239} (\bibinfo {year} {1984})}\BibitemShut {NoStop}%
\bibitem [{\citenamefont {Chou}(1985)}]{CSHY85}%
  \BibitemOpen
  \bibfield  {author} {\bibinfo {author} {\bibfnamefont {K.~C.}\ \bibnamefont
  {Chou}},\ }\href@noop {} {\bibfield  {journal} {\bibinfo  {journal} {Phys.
  Rep. 118 (1985), 1.}\ }\textbf {\bibinfo {volume} {118}},\ \bibinfo {pages}
  {1} (\bibinfo {year} {1985})}\BibitemShut {NoStop}%
\bibitem [{\citenamefont {Rammer}\ and\ \citenamefont {Smith}(1986)}]{RS86}%
  \BibitemOpen
  \bibfield  {author} {\bibinfo {author} {\bibfnamefont {J.}~\bibnamefont
  {Rammer}}\ and\ \bibinfo {author} {\bibfnamefont {H.}~\bibnamefont {Smith}},\
  }\href@noop {} {\bibfield  {journal} {\bibinfo  {journal} {Rev. Mod. Phys. 58
  (1986), 323.}\ }\textbf {\bibinfo {volume} {58}},\ \bibinfo {pages} {323}
  (\bibinfo {year} {1986})}\BibitemShut {NoStop}%
\bibitem [{\citenamefont {Berges}()}]{Berges04}%
  \BibitemOpen
  \bibfield  {author} {\bibinfo {author} {\bibfnamefont {J.}~\bibnamefont
  {Berges}},\ }\href@noop {} {}\bibinfo {note} {Hep-ph0409233.}\BibitemShut
  {Stop}%
\bibitem [{\citenamefont {Haug}\ and\ \citenamefont {Jauho}()}]{HJ98}%
  \BibitemOpen
  \bibfield  {author} {\bibinfo {author} {\bibfnamefont {H.}~\bibnamefont
  {Haug}}\ and\ \bibinfo {author} {\bibfnamefont {A.~P.}\ \bibnamefont
  {Jauho}},\ }\href@noop {} {\bibinfo {title} {{ Quantum Kinetics in Transport
  and Optics of Semiconductors}}},\ \bibinfo {note} {(Springer-Verlag, Berlin,
  1998).}\BibitemShut {Stop}%
\bibitem [{\citenamefont {Rammer}()}]{Rammer07}%
  \BibitemOpen
  \bibfield  {author} {\bibinfo {author} {\bibfnamefont {J.}~\bibnamefont
  {Rammer}},\ }\href@noop {} {\bibinfo {title} {{Quantum Field Theory of
  Nonequilibrium States }}},\ \bibinfo {note} {(Cambridge University Press,
  Cambridge, 2007).}\BibitemShut {Stop}%
\bibitem [{\citenamefont {Kamenev}()}]{Kamenev2}%
  \BibitemOpen
  \bibfield  {author} {\bibinfo {author} {\bibfnamefont {A.}~\bibnamefont
  {Kamenev}},\ }\href@noop {} {\bibinfo {title} {{"Many-body theory of
  non-equilibrium systems"}}},\ \bibinfo {note} {arXiv:cond-mat/0412296
  [cond-mat.dis-nn]}\BibitemShut {NoStop}%
\bibitem [{\citenamefont {Lifshitz}\ and\ \citenamefont {Pitaevskii}()}]{LP}%
  \BibitemOpen
  \bibfield  {author} {\bibinfo {author} {\bibfnamefont {E.~M.}\ \bibnamefont
  {Lifshitz}}\ and\ \bibinfo {author} {\bibfnamefont {L.~P.}\ \bibnamefont
  {Pitaevskii}},\ }\href@noop {} {\bibinfo {title} {{Physical Kinetics }}},\
  \bibinfo {note} {(Pergamon, New York, 1981).}\BibitemShut {Stop}%
\bibitem [{\citenamefont {Mahan}()}]{Mahan}%
  \BibitemOpen
  \bibfield  {author} {\bibinfo {author} {\bibfnamefont {G.~D.}\ \bibnamefont
  {Mahan}},\ }\href@noop {} {\bibinfo {title} {{Many-Particle Physics}}},\
  \bibinfo {note} {(Kluwer Academic/Plenum, New York, 2000).}\BibitemShut
  {Stop}%
\bibitem [{\citenamefont {(Ed.)}()}]{Bonitz00}%
  \BibitemOpen
  \bibfield  {author} {\bibinfo {author} {\bibfnamefont {M.~B.}\ \bibnamefont
  {(Ed.)}},\ }\href@noop {} {\emph {\bibinfo {title} {{Progress in
  Nonequilibrium Green's Functions}}}},\ \bibinfo {note} {(World Scientific,
  Singapore, 2000).}\BibitemShut {Stop}%
\bibitem [{\citenamefont {Landau}(1956)}]{Landau56}%
  \BibitemOpen
  \bibfield  {author} {\bibinfo {author} {\bibfnamefont {L.~D.}\ \bibnamefont
  {Landau}},\ }\href@noop {} {\bibfield  {journal} {\bibinfo  {journal} {Zh.
  Eksp. Teor. Fiz. 30 (1956), 1058 [Sov. Phys. JETP 3 (1957), 920]; Zh. Eksp.
  Teor. Fiz. 32 (1957), 59 [Sov. Phys. JETP 5 (1957), 101].}\ }\textbf
  {\bibinfo {volume} {30}},\ \bibinfo {pages} {1058} (\bibinfo {year}
  {1956})}\BibitemShut {NoStop}%
\bibitem [{\citenamefont {Luttinger}\ and\ \citenamefont {Ward}(1960)}]{LW60}%
  \BibitemOpen
  \bibfield  {author} {\bibinfo {author} {\bibfnamefont {J.~M.}\ \bibnamefont
  {Luttinger}}\ and\ \bibinfo {author} {\bibfnamefont {J.~C.}\ \bibnamefont
  {Ward}},\ }\href@noop {} {\bibfield  {journal} {\bibinfo  {journal} {Phys.
  Rev. 118 (1960), 1417.}\ }\textbf {\bibinfo {volume} {118}},\ \bibinfo
  {pages} {1417} (\bibinfo {year} {1960})}\BibitemShut {NoStop}%
\bibitem [{\citenamefont {Luttinger}(1960)}]{Luttinger60}%
  \BibitemOpen
  \bibfield  {author} {\bibinfo {author} {\bibfnamefont {J.~M.}\ \bibnamefont
  {Luttinger}},\ }\href@noop {} {\bibfield  {journal} {\bibinfo  {journal}
  {Phys. Rev. 119 (1960), 1153.}\ }\textbf {\bibinfo {volume} {119}},\ \bibinfo
  {pages} {1153} (\bibinfo {year} {1960})}\BibitemShut {NoStop}%
\bibitem [{\citenamefont {Cercignani}()}]{Cercignani88}%
  \BibitemOpen
  \bibfield  {author} {\bibinfo {author} {\bibfnamefont {C.}~\bibnamefont
  {Cercignani}},\ }\href@noop {} {\bibinfo {title} {{The Boltzmann Equation and
  Its Applications}}},\ \bibinfo {note} {(Springer-Verlag, New York,
  1988).}\BibitemShut {Stop}%
\bibitem [{\citenamefont {C.~Caroli}\ and\ \citenamefont
  {Saint-James}(1971)}]{CCNS71}%
  \BibitemOpen
  \bibfield  {author} {\bibinfo {author} {\bibfnamefont {P.~N.}\ \bibnamefont
  {C.~Caroli}, \bibfnamefont {R.~Combescot}}\ and\ \bibinfo {author}
  {\bibfnamefont {D.}~\bibnamefont {Saint-James}},\ }\href@noop {} {\bibfield
  {journal} {\bibinfo  {journal} {J. of Phys. C: Solid St. Phys.}\ }\textbf
  {\bibinfo {volume} {4}},\ \bibinfo {pages} {916} (\bibinfo {year}
  {1971})}\BibitemShut {NoStop}%
\bibitem [{\citenamefont {Aronov}\ and\ \citenamefont {Gurevich}(1974)}]{AG75}%
  \BibitemOpen
  \bibfield  {author} {\bibinfo {author} {\bibfnamefont {A.~G.}\ \bibnamefont
  {Aronov}}\ and\ \bibinfo {author} {\bibfnamefont {V.~L.}\ \bibnamefont
  {Gurevich}},\ }\href@noop {} {\bibfield  {journal} {\bibinfo  {journal} {Fiz.
  Tverd. Tela 16 (1974), 2656 [Sov. Phys. Solid State 16 (1975), 1722].}\
  }\textbf {\bibinfo {volume} {16}},\ \bibinfo {pages} {2656} (\bibinfo {year}
  {1974})}\BibitemShut {NoStop}%
\bibitem [{\citenamefont {Larkin}\ and\ \citenamefont
  {Ovchinnikov}(1975)}]{LO75}%
  \BibitemOpen
  \bibfield  {author} {\bibinfo {author} {\bibfnamefont {A.~I.}\ \bibnamefont
  {Larkin}}\ and\ \bibinfo {author} {\bibfnamefont {Y.~B.}\ \bibnamefont
  {Ovchinnikov}},\ }\href@noop {} {\bibfield  {journal} {\bibinfo  {journal}
  {Zh. Eksp. Teor. Fiz. 68 (1975), 1915 [Sov. Phys. JETP 41 (1975), 960].}\
  }\textbf {\bibinfo {volume} {68}},\ \bibinfo {pages} {1915} (\bibinfo {year}
  {1975})}\BibitemShut {NoStop}%
\bibitem [{\citenamefont {Iancu}(2001)}]{CGC}%
  \BibitemOpen
  \bibfield  {author} {\bibinfo {author} {\bibfnamefont {E.}~\bibnamefont
  {Iancu}},\ }\href@noop {} {\bibfield  {journal} {\bibinfo  {journal} {E.
  Iancu, A. Leonidov, and L. D. McLerran, Nucl.Phys.A692, 583 (2001)}\ }
  (\bibinfo {year} {2001})}\BibitemShut {NoStop}%
\bibitem [{\citenamefont {Jensen}(2018)}]{hydr}%
  \BibitemOpen
  \bibfield  {author} {\bibinfo {author} {\bibfnamefont {K.~e.~a.}\
  \bibnamefont {Jensen}},\ }\href@noop {} {\bibinfo {title} {{ “A Panoply of
  Schwinger-Keldysh Transport.”}}} (\bibinfo {year} {2018})\BibitemShut
  {NoStop}%
\bibitem [{\citenamefont {Akhmedov}\ and\ \citenamefont {Popov}(2016)}]{Akh}%
  \BibitemOpen
  \bibfield  {author} {\bibinfo {author} {\bibfnamefont {H.~G.}\ \bibnamefont
  {Akhmedov}, \bibfnamefont {Emil~T.}}\ and\ \bibinfo {author} {\bibfnamefont
  {F.~K.}\ \bibnamefont {Popov}},\ }\href@noop {} {\bibinfo {title}
  {{“Hawking Radiation and Secularly Growing Loop Corrections.”}}}
  (\bibinfo {year} {2016})\BibitemShut {NoStop}%
\bibitem [{\citenamefont {Polkovnikov}(2010)}]{Polkovnikov:2009ys}%
  \BibitemOpen
  \bibfield  {author} {\bibinfo {author} {\bibfnamefont {A.}~\bibnamefont
  {Polkovnikov}},\ }\href@noop {} {\bibfield  {journal} {\bibinfo  {journal}
  {Annals Phys. textbf{325} (2010), 1790-1852 doi:10.1016/j.aop.2010.02.006
  [arXiv:0905.3384 [cond-mat.stat-mech]].}\ ,\ \bibinfo {pages} {1790}}
  (\bibinfo {year} {2010})}\BibitemShut {NoStop}%
\bibitem [{\citenamefont {Zubkov}\ and\ \citenamefont {Wu}(2020)}]{ZW2019}%
  \BibitemOpen
  \bibfield  {author} {\bibinfo {author} {\bibfnamefont {M.~A.}\ \bibnamefont
  {Zubkov}}\ and\ \bibinfo {author} {\bibfnamefont {X.}~\bibnamefont {Wu}},\
  }\href {https://doi.org/10.1016/j.aop.2020.168179} {\bibfield  {journal}
  {\bibinfo  {journal} {Annals of Physics}\ }\textbf {\bibinfo {volume}
  {418}},\ \bibinfo {pages} {168179} (\bibinfo {year} {2020})}\BibitemShut
  {NoStop}%
\bibitem [{\citenamefont {Zhang}\ and\ \citenamefont {Zubkov}(2022)}]{ZZ2022}%
  \BibitemOpen
  \bibfield  {author} {\bibinfo {author} {\bibfnamefont {C.}~\bibnamefont
  {Zhang}}\ and\ \bibinfo {author} {\bibfnamefont {M.}~\bibnamefont {Zubkov}},\
  }\href {https://doi.org/https://doi.org/10.1016/j.aop.2022.169016} {\bibfield
   {journal} {\bibinfo  {journal} {Annals of Physics}\ }\textbf {\bibinfo
  {volume} {444}},\ \bibinfo {pages} {169016} (\bibinfo {year}
  {2022})}\BibitemShut {NoStop}%
\bibitem [{\citenamefont {Khaidukov}\ and\ \citenamefont
  {Zubkov}(2017)}]{Khaidukov2017}%
  \BibitemOpen
  \bibfield  {author} {\bibinfo {author} {\bibfnamefont {Z.~V.}\ \bibnamefont
  {Khaidukov}}\ and\ \bibinfo {author} {\bibfnamefont {M.~A.}\ \bibnamefont
  {Zubkov}},\ }\href {https://doi.org/10.1103/PhysRevD.95.074502} {\bibfield
  {journal} {\bibinfo  {journal} {Phys. Rev. D}\ }\textbf {\bibinfo {volume}
  {95}},\ \bibinfo {pages} {074502} (\bibinfo {year} {2017})},\ \Eprint
  {https://arxiv.org/abs/1701.03368} {arXiv:1701.03368} \BibitemShut {NoStop}%
\bibitem [{\citenamefont {Zubkov}\ and\ \citenamefont
  {Khaidukov}(2017)}]{Zubkov2017}%
  \BibitemOpen
  \bibfield  {author} {\bibinfo {author} {\bibfnamefont {M.~A.}\ \bibnamefont
  {Zubkov}}\ and\ \bibinfo {author} {\bibfnamefont {Z.~V.}\ \bibnamefont
  {Khaidukov}},\ }\href {https://doi.org/10.1134/S0021364017150139} {\bibfield
  {journal} {\bibinfo  {journal} {JETP Lett.}\ }\textbf {\bibinfo {volume}
  {106}},\ \bibinfo {pages} {166} (\bibinfo {year} {2017})},\ \bibinfo {note}
  {[Pisma Zh. Eksp. Teor. Fiz. {\bf 106} no.3, 166]}\BibitemShut {NoStop}%
\bibitem [{\citenamefont {Suleymanov}\ and\ \citenamefont
  {Zubkov}(2020)}]{SuleymanovZubkov2020}%
  \BibitemOpen
  \bibfield  {author} {\bibinfo {author} {\bibfnamefont {M.}~\bibnamefont
  {Suleymanov}}\ and\ \bibinfo {author} {\bibfnamefont {M.~A.}\ \bibnamefont
  {Zubkov}},\ }\href {https://doi.org/10.1103/PhysRevD.102.076019} {\bibfield
  {journal} {\bibinfo  {journal} {Physical Review D}\ }\textbf {\bibinfo
  {volume} {102}},\ \bibinfo {pages} {076019} (\bibinfo {year}
  {2020})}\BibitemShut {NoStop}%
\bibitem [{\citenamefont {Zubkov}\ and\ \citenamefont
  {Abramchuk}(2023{\natexlab{a}})}]{ZA2023}%
  \BibitemOpen
  \bibfield  {author} {\bibinfo {author} {\bibfnamefont {M.}~\bibnamefont
  {Zubkov}}\ and\ \bibinfo {author} {\bibfnamefont {R.~A.}\ \bibnamefont
  {Abramchuk}},\ }\href@noop {} {\bibfield  {journal} {\bibinfo  {journal}
  {Physical Review D}\ }\textbf {\bibinfo {volume} {107}},\ \bibinfo {pages}
  {094021} (\bibinfo {year} {2023}{\natexlab{a}})}\BibitemShut {NoStop}%
\bibitem [{\citenamefont {Zubkov}(2024)}]{Z2024}%
  \BibitemOpen
  \bibfield  {author} {\bibinfo {author} {\bibfnamefont {M.~A.}\ \bibnamefont
  {Zubkov}},\ }\href {https://doi.org/10.1088/1361-648X/ad5d36} {\bibfield
  {journal} {\bibinfo  {journal} {J. Phys. Condens. Matter}\ }\textbf {\bibinfo
  {volume} {36}},\ \bibinfo {pages} {415501} (\bibinfo {year} {2024})},\
  \Eprint {https://arxiv.org/abs/2311.12712} {arXiv:2311.12712
  [cond-mat.mes-hall]} \BibitemShut {NoStop}%
\bibitem [{\citenamefont {Xavier}\ and\ \citenamefont {Zubkov}(2024)}]{XZ2024}%
  \BibitemOpen
  \bibfield  {author} {\bibinfo {author} {\bibfnamefont {P.~D.}\ \bibnamefont
  {Xavier}}\ and\ \bibinfo {author} {\bibfnamefont {M.}~\bibnamefont
  {Zubkov}},\ }\href@noop {} {\bibfield  {journal} {\bibinfo  {journal} {arXiv
  preprint arXiv:2410.06952, to appear in Physical Review D}\ } (\bibinfo
  {year} {2024})}\BibitemShut {NoStop}%
\bibitem [{\citenamefont {Shitade}(2017)}]{Shitade}%
  \BibitemOpen
  \bibfield  {author} {\bibinfo {author} {\bibfnamefont {A.}~\bibnamefont
  {Shitade}},\ }\href {https://doi.org/10.7566/JPSJ.86.054601} {\bibfield
  {journal} {\bibinfo  {journal} {Journal of the Physical Society of Japan}\
  }\textbf {\bibinfo {volume} {86}},\ \bibinfo {pages} {054601} (\bibinfo
  {year} {2017})},\ \Eprint
  {https://arxiv.org/abs/https://doi.org/10.7566/JPSJ.86.054601}
  {https://doi.org/10.7566/JPSJ.86.054601} \BibitemShut {NoStop}%
\bibitem [{\citenamefont {Lux}\ \emph {et~al.}(2020)\citenamefont {Lux},
  \citenamefont {Freimuth}, \citenamefont {Bl\"ugel},\ and\ \citenamefont
  {Mokrousov}}]{Mokrousov}%
  \BibitemOpen
  \bibfield  {author} {\bibinfo {author} {\bibfnamefont {F.~R.}\ \bibnamefont
  {Lux}}, \bibinfo {author} {\bibfnamefont {F.}~\bibnamefont {Freimuth}},
  \bibinfo {author} {\bibfnamefont {S.}~\bibnamefont {Bl\"ugel}},\ and\
  \bibinfo {author} {\bibfnamefont {Y.}~\bibnamefont {Mokrousov}},\ }\href
  {https://doi.org/10.1103/PhysRevLett.124.096602} {\bibfield  {journal}
  {\bibinfo  {journal} {Phys. Rev. Lett.}\ }\textbf {\bibinfo {volume} {124}},\
  \bibinfo {pages} {096602} (\bibinfo {year} {2020})}\BibitemShut {NoStop}%
\bibitem [{ZUM(1984)}]{ZUMINO1984477}%
  \BibitemOpen
  \href {https://doi.org/https://doi.org/10.1016/0550-3213(84)90259-1}
  {\bibfield  {journal} {\bibinfo  {journal} {Nuclear Physics B}\ }\textbf
  {\bibinfo {volume} {239}},\ \bibinfo {pages} {477} (\bibinfo {year}
  {1984})}\BibitemShut {NoStop}%
\bibitem [{\citenamefont {Thouless}(1983)}]{PhysRevB.27.6083}%
  \BibitemOpen
  \bibfield  {author} {\bibinfo {author} {\bibfnamefont {D.~J.}\ \bibnamefont
  {Thouless}},\ }\href {https://doi.org/10.1103/PhysRevB.27.6083} {\bibfield
  {journal} {\bibinfo  {journal} {Phys. Rev. B}\ }\textbf {\bibinfo {volume}
  {27}},\ \bibinfo {pages} {6083} (\bibinfo {year} {1983})}\BibitemShut
  {NoStop}%
\bibitem [{\citenamefont {Niu}(1990)}]{PhysRevLett.64.1812}%
  \BibitemOpen
  \bibfield  {author} {\bibinfo {author} {\bibfnamefont {Q.}~\bibnamefont
  {Niu}},\ }\href {https://doi.org/10.1103/PhysRevLett.64.1812} {\bibfield
  {journal} {\bibinfo  {journal} {Phys. Rev. Lett.}\ }\textbf {\bibinfo
  {volume} {64}},\ \bibinfo {pages} {1812} (\bibinfo {year}
  {1990})}\BibitemShut {NoStop}%
\bibitem [{\citenamefont {Lohse}\ \emph {et~al.}(2016)\citenamefont {Lohse},
  \citenamefont {Schweizer}, \citenamefont {Zilberberg}, \citenamefont
  {Aidelsburger},\ and\ \citenamefont {Bloch}}]{Lohse2016}%
  \BibitemOpen
  \bibfield  {author} {\bibinfo {author} {\bibfnamefont {M.}~\bibnamefont
  {Lohse}}, \bibinfo {author} {\bibfnamefont {C.}~\bibnamefont {Schweizer}},
  \bibinfo {author} {\bibfnamefont {O.}~\bibnamefont {Zilberberg}}, \bibinfo
  {author} {\bibfnamefont {M.}~\bibnamefont {Aidelsburger}},\ and\ \bibinfo
  {author} {\bibfnamefont {I.}~\bibnamefont {Bloch}},\ }\href
  {https://doi.org/10.1038/nphys3584} {\bibfield  {journal} {\bibinfo
  {journal} {Nature Physics}\ }\textbf {\bibinfo {volume} {12}},\ \bibinfo
  {pages} {350} (\bibinfo {year} {2016})}\BibitemShut {NoStop}%
\bibitem [{\citenamefont {Nakajima}\ \emph {et~al.}(2016)\citenamefont
  {Nakajima}, \citenamefont {Tomita}, \citenamefont {Taie}, \citenamefont
  {Ichinose}, \citenamefont {Ozawa}, \citenamefont {Wang}, \citenamefont
  {Troyer},\ and\ \citenamefont {Takahashi}}]{Nakajima2016}%
  \BibitemOpen
  \bibfield  {author} {\bibinfo {author} {\bibfnamefont {S.}~\bibnamefont
  {Nakajima}}, \bibinfo {author} {\bibfnamefont {T.}~\bibnamefont {Tomita}},
  \bibinfo {author} {\bibfnamefont {S.}~\bibnamefont {Taie}}, \bibinfo {author}
  {\bibfnamefont {T.}~\bibnamefont {Ichinose}}, \bibinfo {author}
  {\bibfnamefont {H.}~\bibnamefont {Ozawa}}, \bibinfo {author} {\bibfnamefont
  {L.}~\bibnamefont {Wang}}, \bibinfo {author} {\bibfnamefont {M.}~\bibnamefont
  {Troyer}},\ and\ \bibinfo {author} {\bibfnamefont {Y.}~\bibnamefont
  {Takahashi}},\ }\href {https://doi.org/10.1038/nphys3622} {\bibfield
  {journal} {\bibinfo  {journal} {Nature Physics}\ }\textbf {\bibinfo {volume}
  {12}},\ \bibinfo {pages} {296} (\bibinfo {year} {2016})}\BibitemShut
  {NoStop}%
\bibitem [{\citenamefont {Bevan}\ \emph {et~al.}(1997)\citenamefont {Bevan},
  \citenamefont {Manninen}, \citenamefont {Cook}, \citenamefont {Hook},
  \citenamefont {Hall}, \citenamefont {Vachaspati},\ and\ \citenamefont
  {Volovik}}]{bevan1997momentum}%
  \BibitemOpen
  \bibfield  {author} {\bibinfo {author} {\bibfnamefont {T.}~\bibnamefont
  {Bevan}}, \bibinfo {author} {\bibfnamefont {A.}~\bibnamefont {Manninen}},
  \bibinfo {author} {\bibfnamefont {J.}~\bibnamefont {Cook}}, \bibinfo {author}
  {\bibfnamefont {J.}~\bibnamefont {Hook}}, \bibinfo {author} {\bibfnamefont
  {H.}~\bibnamefont {Hall}}, \bibinfo {author} {\bibfnamefont {T.}~\bibnamefont
  {Vachaspati}},\ and\ \bibinfo {author} {\bibfnamefont {G.}~\bibnamefont
  {Volovik}},\ }\href@noop {} {\bibfield  {journal} {\bibinfo  {journal}
  {Nature}\ }\textbf {\bibinfo {volume} {386}},\ \bibinfo {pages} {689}
  (\bibinfo {year} {1997})}\BibitemShut {NoStop}%
\bibitem [{\citenamefont {Drude}(1900)}]{Drude1900}%
  \BibitemOpen
  \bibfield  {author} {\bibinfo {author} {\bibfnamefont {P.}~\bibnamefont
  {Drude}},\ }\href {https://doi.org/https://doi.org/10.1002/andp.19003060312}
  {\bibfield  {journal} {\bibinfo  {journal} {Annalen der Physik}\ }\textbf
  {\bibinfo {volume} {306}},\ \bibinfo {pages} {566} (\bibinfo {year}
  {1900})}\BibitemShut {NoStop}%
\bibitem [{\citenamefont {Zhang}\ and\ \citenamefont
  {Zhou}(2017)}]{PhysRevA.95.061601}%
  \BibitemOpen
  \bibfield  {author} {\bibinfo {author} {\bibfnamefont {S.-L.}\ \bibnamefont
  {Zhang}}\ and\ \bibinfo {author} {\bibfnamefont {Q.}~\bibnamefont {Zhou}},\
  }\href {https://doi.org/10.1103/PhysRevA.95.061601} {\bibfield  {journal}
  {\bibinfo  {journal} {Phys. Rev. A}\ }\textbf {\bibinfo {volume} {95}},\
  \bibinfo {pages} {061601} (\bibinfo {year} {2017})}\BibitemShut {NoStop}%
\bibitem [{\citenamefont {Gusynin}\ \emph {et~al.}(1999)\citenamefont
  {Gusynin}, \citenamefont {Miransky},\ and\ \citenamefont
  {Shovkovy}}]{Gusynin1999pq}%
  \BibitemOpen
  \bibfield  {author} {\bibinfo {author} {\bibfnamefont {V.~P.}\ \bibnamefont
  {Gusynin}}, \bibinfo {author} {\bibfnamefont {V.~A.}\ \bibnamefont
  {Miransky}},\ and\ \bibinfo {author} {\bibfnamefont {I.~A.}\ \bibnamefont
  {Shovkovy}},\ }\href {https://doi.org/10.1016/S0550-3213(99)00573-8}
  {\bibfield  {journal} {\bibinfo  {journal} {Nucl. Phys. B}\ }\textbf
  {\bibinfo {volume} {563}},\ \bibinfo {pages} {361} (\bibinfo {year}
  {1999})},\ \Eprint {https://arxiv.org/abs/hep-ph/9908320}
  {arXiv:hep-ph/9908320} \BibitemShut {NoStop}%
\bibitem [{\citenamefont {Abramchuk}\ and\ \citenamefont
  {Zubkov}(2024)}]{abramchuk2024magnetoconductivity}%
  \BibitemOpen
  \bibfield  {author} {\bibinfo {author} {\bibfnamefont {R.}~\bibnamefont
  {Abramchuk}}\ and\ \bibinfo {author} {\bibfnamefont {M.}~\bibnamefont
  {Zubkov}},\ }\href@noop {} {\bibinfo {title} {Magnetoconductivity of dirac
  semimetals and chiral magnetic effect from keldysh technique}} (\bibinfo
  {year} {2024}),\ \bibinfo {note} {arXiv preprint
  arXiv:2409.14941}\BibitemShut {NoStop}%
\bibitem [{\citenamefont {Abramchuk}(2026)}]{ABRAMCHUK2026113374}%
  \BibitemOpen
  \bibfield  {author} {\bibinfo {author} {\bibfnamefont {R.~A.}\ \bibnamefont
  {Abramchuk}},\ }\href
  {https://doi.org/https://doi.org/10.1016/j.jpcs.2025.113374} {\bibfield
  {journal} {\bibinfo  {journal} {Journal of Physics and Chemistry of Solids}\
  }\textbf {\bibinfo {volume} {210}},\ \bibinfo {pages} {113374} (\bibinfo
  {year} {2026})}\BibitemShut {NoStop}%
\bibitem [{\citenamefont {Li}\ \emph {et~al.}(2016)\citenamefont {Li},
  \citenamefont {Kharzeev}, \citenamefont {Zhang}, \citenamefont {Huang},
  \citenamefont {Pletikosic}, \citenamefont {Fedorov}, \citenamefont {Zhong},
  \citenamefont {Schneeloch}, \citenamefont {Gu},\ and\ \citenamefont
  {Valla}}]{CMEZrTe5}%
  \BibitemOpen
  \bibfield  {author} {\bibinfo {author} {\bibfnamefont {Q.}~\bibnamefont
  {Li}}, \bibinfo {author} {\bibfnamefont {D.~E.}\ \bibnamefont {Kharzeev}},
  \bibinfo {author} {\bibfnamefont {C.}~\bibnamefont {Zhang}}, \bibinfo
  {author} {\bibfnamefont {Y.}~\bibnamefont {Huang}}, \bibinfo {author}
  {\bibfnamefont {I.}~\bibnamefont {Pletikosic}}, \bibinfo {author}
  {\bibfnamefont {A.~V.}\ \bibnamefont {Fedorov}}, \bibinfo {author}
  {\bibfnamefont {R.~D.}\ \bibnamefont {Zhong}}, \bibinfo {author}
  {\bibfnamefont {J.~A.}\ \bibnamefont {Schneeloch}}, \bibinfo {author}
  {\bibfnamefont {G.~D.}\ \bibnamefont {Gu}},\ and\ \bibinfo {author}
  {\bibfnamefont {T.}~\bibnamefont {Valla}},\ }\href
  {https://doi.org/10.1038/nphys3648} {\bibfield  {journal} {\bibinfo
  {journal} {Nature Phys.}\ }\textbf {\bibinfo {volume} {12}},\ \bibinfo
  {pages} {550} (\bibinfo {year} {2016})},\ \Eprint
  {https://arxiv.org/abs/1412.6543} {arXiv:1412.6543 [cond-mat.str-el]}
  \BibitemShut {NoStop}%
\bibitem [{\citenamefont {Kaushik}\ and\ \citenamefont
  {Kharzeev}(2017)}]{Kharzeev2017}%
  \BibitemOpen
  \bibfield  {author} {\bibinfo {author} {\bibfnamefont {S.}~\bibnamefont
  {Kaushik}}\ and\ \bibinfo {author} {\bibfnamefont {D.~E.}\ \bibnamefont
  {Kharzeev}},\ }\href {https://doi.org/10.1103/PhysRevB.95.235136} {\bibfield
  {journal} {\bibinfo  {journal} {Phys. Rev. B}\ }\textbf {\bibinfo {volume}
  {95}},\ \bibinfo {pages} {235136} (\bibinfo {year} {2017})},\ \Eprint
  {https://arxiv.org/abs/1703.05865} {arXiv:1703.05865 [cond-mat.mes-hall]}
  \BibitemShut {NoStop}%
\bibitem [{\citenamefont {Meier}\ \emph {et~al.}(2016)\citenamefont {Meier},
  \citenamefont {An},\ and\ \citenamefont {Gadway}}]{meier2016observation}%
  \BibitemOpen
  \bibfield  {author} {\bibinfo {author} {\bibfnamefont {E.~J.}\ \bibnamefont
  {Meier}}, \bibinfo {author} {\bibfnamefont {F.~A.}\ \bibnamefont {An}},\ and\
  \bibinfo {author} {\bibfnamefont {B.}~\bibnamefont {Gadway}},\ }\href@noop {}
  {\bibfield  {journal} {\bibinfo  {journal} {Nature communications}\ }\textbf
  {\bibinfo {volume} {7}},\ \bibinfo {pages} {13986} (\bibinfo {year}
  {2016})}\BibitemShut {NoStop}%
\bibitem [{\citenamefont {Qin}\ \emph {et~al.}(2023{\natexlab{b}})\citenamefont
  {Qin}, \citenamefont {Xu}, \citenamefont {Ning},\ and\ \citenamefont
  {Wang}}]{qin2023one}%
  \BibitemOpen
  \bibfield  {author} {\bibinfo {author} {\bibfnamefont {Z.}~\bibnamefont
  {Qin}}, \bibinfo {author} {\bibfnamefont {D.-H.}\ \bibnamefont {Xu}},
  \bibinfo {author} {\bibfnamefont {Z.}~\bibnamefont {Ning}},\ and\ \bibinfo
  {author} {\bibfnamefont {R.}~\bibnamefont {Wang}},\ }\href@noop {} {\bibfield
   {journal} {\bibinfo  {journal} {Physical Review B}\ }\textbf {\bibinfo
  {volume} {108}},\ \bibinfo {pages} {195103} (\bibinfo {year}
  {2023}{\natexlab{b}})}\BibitemShut {NoStop}%
\bibitem [{\citenamefont {Banerjee}\ \emph
  {et~al.}(2022{\natexlab{b}})\citenamefont {Banerjee}, \citenamefont
  {Lewkowicz},\ and\ \citenamefont {Zubkov}}]{banerjee2022chiral}%
  \BibitemOpen
  \bibfield  {author} {\bibinfo {author} {\bibfnamefont {C.}~\bibnamefont
  {Banerjee}}, \bibinfo {author} {\bibfnamefont {M.}~\bibnamefont
  {Lewkowicz}},\ and\ \bibinfo {author} {\bibfnamefont {M.~A.}\ \bibnamefont
  {Zubkov}},\ }\href@noop {} {\bibfield  {journal} {\bibinfo  {journal}
  {Physical Review D}\ }\textbf {\bibinfo {volume} {106}},\ \bibinfo {pages}
  {074508} (\bibinfo {year} {2022}{\natexlab{b}})}\BibitemShut {NoStop}%
\bibitem [{\citenamefont {Zubkov}(2017)}]{zubkov2017topology}%
  \BibitemOpen
  \bibfield  {author} {\bibinfo {author} {\bibfnamefont {M.}~\bibnamefont
  {Zubkov}},\ }\href@noop {} {\bibfield  {journal} {\bibinfo  {journal} {JETP
  Letters}\ }\textbf {\bibinfo {volume} {106}},\ \bibinfo {pages} {172}
  (\bibinfo {year} {2017})}\BibitemShut {NoStop}%
\bibitem [{\citenamefont {Volovik}(2003)}]{Volovik2003a}%
  \BibitemOpen
  \bibfield  {author} {\bibinfo {author} {\bibfnamefont {G.~E.}\ \bibnamefont
  {Volovik}},\ }\href@noop {} {\emph {\bibinfo {title} {{The Universe in a
  Helium Droplet}}}}\ (\bibinfo  {publisher} {Clarendon Press},\ \bibinfo
  {address} {Oxford},\ \bibinfo {year} {2003})\BibitemShut {NoStop}%
\bibitem [{\citenamefont {Chernodub}\ and\ \citenamefont
  {Zubkov}(2017)}]{chernodub2017scale}%
  \BibitemOpen
  \bibfield  {author} {\bibinfo {author} {\bibfnamefont {M.~N.}\ \bibnamefont
  {Chernodub}}\ and\ \bibinfo {author} {\bibfnamefont {M.}~\bibnamefont
  {Zubkov}},\ }\href@noop {} {\bibfield  {journal} {\bibinfo  {journal}
  {Physical Review D}\ }\textbf {\bibinfo {volume} {96}},\ \bibinfo {pages}
  {056006} (\bibinfo {year} {2017})}\BibitemShut {NoStop}%
\bibitem [{\citenamefont {Zhang}\ and\ \citenamefont
  {Zubkov}(2020)}]{zhang2020influence}%
  \BibitemOpen
  \bibfield  {author} {\bibinfo {author} {\bibfnamefont {C.}~\bibnamefont
  {Zhang}}\ and\ \bibinfo {author} {\bibfnamefont {M.}~\bibnamefont {Zubkov}},\
  }\href@noop {} {\bibfield  {journal} {\bibinfo  {journal} {Journal of Physics
  A: Mathematical and Theoretical}\ }\textbf {\bibinfo {volume} {53}},\
  \bibinfo {pages} {195002} (\bibinfo {year} {2020})}\BibitemShut {NoStop}%
\bibitem [{\citenamefont {Zubkov}\ and\ \citenamefont
  {Abramchuk}(2023{\natexlab{b}})}]{zubkov2023effect}%
  \BibitemOpen
  \bibfield  {author} {\bibinfo {author} {\bibfnamefont {M.}~\bibnamefont
  {Zubkov}}\ and\ \bibinfo {author} {\bibfnamefont {R.}~\bibnamefont
  {Abramchuk}},\ }\href@noop {} {\bibfield  {journal} {\bibinfo  {journal}
  {Physical Review D}\ }\textbf {\bibinfo {volume} {107}},\ \bibinfo {pages}
  {094021} (\bibinfo {year} {2023}{\natexlab{b}})}\BibitemShut {NoStop}%
\bibitem [{\citenamefont {Zubkov}(2018)}]{zubkov2018momentum}%
  \BibitemOpen
  \bibfield  {author} {\bibinfo {author} {\bibfnamefont {M.}~\bibnamefont
  {Zubkov}},\ }\href@noop {} {\bibfield  {journal} {\bibinfo  {journal} {Annals
  of Physics}\ }\textbf {\bibinfo {volume} {393}},\ \bibinfo {pages} {264}
  (\bibinfo {year} {2018})}\BibitemShut {NoStop}%
\bibitem [{\citenamefont {Zubkov}\ and\ \citenamefont
  {Volovik}(2012)}]{zubkov2012momentum}%
  \BibitemOpen
  \bibfield  {author} {\bibinfo {author} {\bibfnamefont {M.}~\bibnamefont
  {Zubkov}}\ and\ \bibinfo {author} {\bibfnamefont {G.}~\bibnamefont
  {Volovik}},\ }\href@noop {} {\bibfield  {journal} {\bibinfo  {journal}
  {Nuclear Physics B}\ }\textbf {\bibinfo {volume} {860}},\ \bibinfo {pages}
  {295} (\bibinfo {year} {2012})}\BibitemShut {NoStop}%
\bibitem [{\citenamefont {Volovik}\ and\ \citenamefont
  {Zubkov}(2017)}]{volovik2017standard}%
  \BibitemOpen
  \bibfield  {author} {\bibinfo {author} {\bibfnamefont {G.}~\bibnamefont
  {Volovik}}\ and\ \bibinfo {author} {\bibfnamefont {M.}~\bibnamefont
  {Zubkov}},\ }\href@noop {} {\bibfield  {journal} {\bibinfo  {journal} {New
  Journal of Physics}\ }\textbf {\bibinfo {volume} {19}},\ \bibinfo {pages}
  {015009} (\bibinfo {year} {2017})}\BibitemShut {NoStop}%
\bibitem [{\citenamefont {Volovik}\ and\ \citenamefont
  {Zubkov}(2013)}]{volovik2013nambu}%
  \BibitemOpen
  \bibfield  {author} {\bibinfo {author} {\bibfnamefont {G.~E.}\ \bibnamefont
  {Volovik}}\ and\ \bibinfo {author} {\bibfnamefont {M.}~\bibnamefont
  {Zubkov}},\ }\href@noop {} {\bibfield  {journal} {\bibinfo  {journal} {JETP
  letters}\ }\textbf {\bibinfo {volume} {97}},\ \bibinfo {pages} {301}
  (\bibinfo {year} {2013})}\BibitemShut {NoStop}%
\end{thebibliography}%

\end{document}